\documentclass[sigconf]{acmart}

\usepackage{graphicx}
\usepackage{amsmath}

\usepackage{algorithm}
\usepackage{algorithmic}

\usepackage{amsmath}
\usepackage{amsbsy}

\usepackage{enumitem}

\usepackage{booktabs}
\usepackage{multirow}

\usepackage{colortbl}

\usepackage{subcaption}

\usepackage{multicol}

\usepackage{multirow}
\usepackage{array}

\usepackage{mdframed}
\usepackage{lipsum}  % This package provides filler text.
\usepackage{geometry}
\usepackage{tcolorbox}

\usepackage{float}
\usepackage{hyperref} % chaolianjie

\AtBeginDocument{%
  }

\copyrightyear{2025}
\acmYear{2025}
\setcopyright{acmlicensed}\acmConference[WWW '25]{Proceedings of the ACM Web Conference 2025}{April 28-May 2, 2025}{Sydney, NSW, Australia}
\acmBooktitle{Proceedings of the ACM Web Conference 2025 (WWW '25), April 28-May 2, 2025, Sydney, NSW, Australia}
\acmDOI{10.1145/3696410.3714863}
\acmISBN{979-8-4007-1274-6/25/04}

\begin{document}

%%
%% The "title" command has an optional parameter,
%% allowing the author to define a "short title" to be used in page headers.
\title{TourRank: Utilizing Large Language Models for Documents Ranking with a Tournament-Inspired Strategy}

\author{Yiqun Chen}
\affiliation{%
  \institution{Renmin University of China}
  \city{Beijing}
  \country{China}}
\email{chenyiqun990321@ruc.edu.cn}

\author{Qi Liu}
\affiliation{%
  \institution{Renmin University of China}
  \city{Beijing}
  \country{China}}
\email{liuqi_67@ruc.edu.cn}

\author{Yi Zhang}
\affiliation{%
  \institution{Baidu Inc.}
  \city{Beijing}
  \country{China}}
\email{zhangyi75@baidu.com}

\author{Weiwei Sun}
\affiliation{%
  \institution{Carnegie Mellon University}
  \city{Pittsburgh}
  \country{USA}}
\email{sunnweiwei@gmail.com}

\author{Xinyu Ma}
\affiliation{%
  \institution{Baidu Inc.}
  \city{Beijing}
  \country{China}}
\email{xinyuma2016@gmail.com}

\author{Wei Yang}
\affiliation{%
  \institution{University of Southern California}
  \city{Los Angeles}
  \country{USA}}
\email{weiyangvia@gmail.com}

\author{Daiting Shi}
\affiliation{%
  \institution{Baidu Inc.}
  \city{Beijing}
  \country{China}}
\email{shidaiting01@baidu.com}

\author{Jiaxin Mao}
\authornote{Jiaxin Mao and Dawei Yin are the corresponding authors.}
\affiliation{%
  % \institution{Gaoling School of Artificial Intelligence, Renmin University of China}
  \institution{Renmin University of China}
  \city{Beijing}
  \country{China}}
\email{maojiaxin@gmail.com}

\author{Dawei Yin}
\authornotemark[1]
\affiliation{%
  \institution{Baidu Inc.}
  \city{Beijing}
  \country{China}}
\email{yindawei@acm.org}

%%
%% By default, the full list of authors will be used in the page
%% headers. Often, this list is too long, and will overlap
%% other information printed in the page headers. This command allows
%% the author to define a more concise list
%% of authors' names for this purpose.
% \renewcommand{\shortauthors}{Trovato et al.}
\renewcommand{\shortauthors}{Yiqun Chen et al.}

%%
%% The abstract is a short summary of the work to be presented in the
%% article.
\begin{abstract}
Large Language Models (LLMs) are increasingly employed in zero-shot documents ranking, yielding commendable results. However, several significant challenges still persist in LLMs for ranking: (1) LLMs are constrained by limited input length, precluding them from processing a large number of documents simultaneously; (2) The output document sequence is influenced by the input order of documents, resulting in inconsistent ranking outcomes; (3) Achieving a balance between cost and ranking performance is challenging. To tackle these issues, we introduce a novel documents ranking method called TourRank \footnote{The code of TourRank can be seen on \url{https://github.com/chenyiqun/TourRank}.}, which is inspired by the sport tournaments, such as FIFA World Cup. Specifically, we 1) overcome the limitation in input length and reduce the ranking latency by incorporating a multi-stage grouping strategy similar to the parallel group stage of sport tournaments; 2) improve the ranking performance and robustness to input orders by using a points system to ensemble multiple ranking results. We test TourRank with different LLMs on the TREC DL datasets and the BEIR benchmark. The experimental results demonstrate that TourRank delivers state-of-the-art performance at a modest cost.
\end{abstract}

%%
%% The code below is generated by the tool at http://dl.acm.org/ccs.cfm.
%% Please copy and paste the code instead of the example below.
%%

\begin{CCSXML}
<ccs2012>
   <concept>
       <concept_id>10002951.10003317.10003338.10003341</concept_id>
       <concept_desc>Information systems~Language models</concept_desc>
       <concept_significance>500</concept_significance>
       </concept>
 </ccs2012>
\end{CCSXML}

\ccsdesc[500]{Information systems~Language models}

% \begin{CCSXML}
% <ccs2012>
%  <concept>
%   <concept_id>00000000.0000000.0000000</concept_id>
%   <concept_desc>Do Not Use This Code, Generate the Correct Terms for Your Paper</concept_desc>
%   <concept_significance>500</concept_significance>
%  </concept>
%  <concept>
%   <concept_id>00000000.00000000.00000000</concept_id>
%   <concept_desc>Do Not Use This Code, Generate the Correct Terms for Your Paper</concept_desc>
%   <concept_significance>300</concept_significance>
%  </concept>
%  <concept>
%   <concept_id>00000000.00000000.00000000</concept_id>
%   <concept_desc>Do Not Use This Code, Generate the Correct Terms for Your Paper</concept_desc>
%   <concept_significance>100</concept_significance>
%  </concept>
%  <concept>
%   <concept_id>00000000.00000000.00000000</concept_id>
%   <concept_desc>Do Not Use This Code, Generate the Correct Terms for Your Paper</concept_desc>
%   <concept_significance>100</concept_significance>
%  </concept>
% </ccs2012>
% \end{CCSXML}

% \ccsdesc[500]{Do Not Use This Code~Generate the Correct Terms for Your Paper}
% \ccsdesc[300]{Do Not Use This Code~Generate the Correct Terms for Your Paper}
% \ccsdesc{Do Not Use This Code~Generate the Correct Terms for Your Paper}
% \ccsdesc[100]{Do Not Use This Code~Generate the Correct Terms for Your Paper}

%%
%% Keywords. The author(s) should pick words that accurately describe
%% the work being presented. Separate the keywords with commas.
\keywords{Large Language Model for Search; Zero-Shot Ranking}
%% A "teaser" image appears between the author and affiliation
%% information and the body of the document, and typically spans the
%% page.

% \received{20 February 2007}
% \received[revised]{12 March 2009}
% \received[accepted]{5 June 2009}

%%
%% This command processes the author and affiliation and title
%% information and builds the first part of the formatted document.
\maketitle

\section{Introduction}

Recently, Large Language Models (LLMs) have demonstrated great potential in numerous Natural Language Processing (NLP) tasks, especially under the zero-shot settings. Researchers and practitioners have also tried to leverage LLMs document ranking \cite{zhu2023large}, a core task in information retrieval, under the zero-shot settings. 
Most of the existing LLM-based document ranking methods can be divided into three categories: the \textit{Pointwise} approach that prompts LLMs to independetly assess the relevance of each candidate document~\cite{sachan2022improving, liang2022holistic, zhuang2023beyond, guo2024generating, sun2023instruction}; the \textit{Pairwise} approach that use LLMs to compare each document against all the other documents~\cite{qin2023large}; and the \textit{Listwise} approach that instruct LLMs to generate a ranked list of document labels according to their relevance to the query~\cite{sun2023chatgpt, ma2023zero, pradeep2023rankvicuna, pradeep2023rankzephyr, zhuang2023setwise}.

% \footnote{See Appendix \ref{LLMs Approaches} for a more detailed literature review of the pointwise, pairwise, and listwise approaches for LLM-based document ranking.}

While these three approaches lead to different trade-offs between effectiveness and efficiency, the listwise approach, such as RankGPT \cite{sun2023chatgpt}, is considered as the preferred prompting strategy for the LLM-based zero-shot document ranking task.
Unlike the pointwise approach, the listwise approach considers multiple documents simultaneously and thus yields better effectiveness in ranking. Meanwhile, listwise ranking eludes the quadratic growing cost of comparing every pair of documents in the candidate list, resulting in improved efficiency than the pairwise approach.

Although the listwise approaches achieve a good trade-off between effectiveness and efficiency and thus are considered preferred prompting strategies for LLM-based document ranking, they also face certain challenges: 
(1) The maximum context length of LLMs limits the number of documents that can be compared in a single prompt;
(2) The listwise generation process can not run in parallel, which makes it hard to return the final ranking list under a tight time constraint.
(3) The ranking results are highly dependent on the initial order of the candidate documents in the input prompt.
% (1) The limited input length of LLMs restricts the number of documents that can be compared simultaneously; 
% (2) The ranking outcome is highly dependent on the initial order of the candidate documents; 
% (3) The multiple steps contained in many listwise methods can not be run in parallel, which makes it hard to return the final ranking under a tight time constraint. 

To address these challenges, we need to develop a prompting strategy for LLM-based document ranking that can: (Requirement 1) establish a global ranking for about 100 candidate documents through multiple local comparisons of 2 to 10 documents in a single prompt; (Requirement 2) parallelize multiple LLM inferences to minimize the overall ranking time; and (Requirement 3) effectively leverages the initial order of candidate documents set by the first-stage retrieval model without becoming overly dependent on it.
% In order to overcome the above challenges, the following capabilities are required in LLM-based document ranking methods: (1) The method should have the ability to divide all candidates into multiple parts to meet the limited input length, and then integrate the multiple parts into a whole; (2) The method can make full use of the initial permutation but should not rely too much on it; (3) Some parts of the method should be able to run in parallel to give the ranking list in the shortest possible time. 

\begin{figure}[t]
  \centering
  \includegraphics[width=0.45\textwidth]{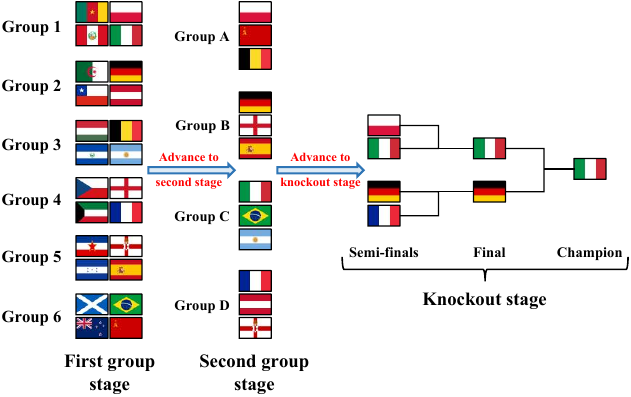}
  \vspace{-2.5mm}
  \caption{The 1982 FIFA World Cup. \\ \small{In the first group stage, 24 teams were divided into six groups, and the top 2 out of 4 teams in each group qualified. In the second group stage, 12 teams were divided into 4 groups, and only the top 1 out of 3 teams in each group advanced. In knockout stages, only the winner in each two-team match progressed to the next stage.}}
  \label{world_cup}
  \vspace{-5mm}
\end{figure}

Interestingly, we find that using LLMs and prompts to rank documents for a query can be analogous to ranking teams or athletes in a sports tournament, as the design of a sports tournament has similar requirements.
A tournament in sports is a structured competition involving multiple teams or individual competitors who compete against each other in a series of matches or games,  with the goal of determining a champion or ranking the participants. 
Figure \ref{world_cup} shows the format and results of an example tournament, the 1982 FIFA World Cup. 
The tournament consists of two group stages and two knockout stages (i.e., the semi-finals and the final). 
Analogous to Requirement 1, each group in the group stages and each two-team match in the knockout stages served as a local comparison; the results of these local matches determined which teams could advance to the next stage and their final rankings in the tournament.
To expedite the ranking process, the World Cup organized multiple parallel matches across different groups. This parallelization allowed the tournament to progress efficiently and fit into a tight 4-week schedule, which meets Requirement 2.
Regarding Requirement 3, the initial groupings were based on seeding and previous performance, providing an initial order of teams. However, the tournament did not solely rely on these seedings; each team’s performance in the group stage and subsequent rounds determined their advancement and final rankings.
% Coincidentally, the goal of a tournament \footnote{The tournament in sports is a structured competition involving candidate teams or competitors who compete against each other in a series of matches or games, with the goal of determining a champion or ranking the participants.} in sports is to rank the participants which is very similar to the goal of ranking the candidate documents. For example, the 1982 FIFA World Cup shown in Figure \ref{world_cup} contains multiple stages, and multiple matches are carried out in parallel in each stage. At each stage, each group selects one or more countries to advance to the next stage. After this tournament (FIFA), the champion is selected and all candidate countries are ranked efficiently and accurately. 

Therefore, inspired by the tournament mechanism, we propose a new zero-shot document ranking method called \textbf{TourRank}, which can fulfill the three requirements and mitigate the challenges in existing methods. In TourRank, we regard each candidate document as a participant in a multi-stage tournament. 
In each stage, we group the candidate documents and prompt the LLM to select the most relevant documents in each group to advance to the next stage. The LLM inferences across different groups in a single stage can be parallelized. 
We also design a grouping strategy, similar to the seeding strategy in sports tournaments, to make use of the initial document order provided by the first-stage retrieval model in ranking. 
In addition, to further improve the effectiveness and robustness, we design a points system to assign different points to each candidate document based on its ranking in each round tournament and perform multiple rounds of tournament. In this way, we can ensemble the results in each round of tournament into a single ranking list based on the final accumulated points in descending order.

To demonstrate the effectiveness of our approach, We test TourRank and baselines on the TREC DL 19 \cite{craswell2020overview}, TREC DL 20 datasets \cite{craswell2021overview}, and 8 datasets from BEIR benchmark \cite{thakur2021beir}. TourRank achieves state-of-the-art performance on the TREC DL datasets and the most tasks of BEIR benchmark, and achieves a good balance between performance and resource consumption. Experiments on different retriever and initial orderings demonstrate the robustness of TourRank ranking. We further evaluate TourRank using a range of large language models (LLMs), including gpt-3.5-turbo, gpt-4-turbo, gpt-4o-mini via OpenAI’s API, as well as several open-source models, such as Mistral-7B-Instruct-v0.2 \cite{jiang2023mistral}, Llama-3-8B-Instruct \cite{Meta2024} and vicuna-13b-v1.5 \cite{zheng2023judging}. The results suggest that TourRank consistently outperform some existing listwise ranking approaches.

% 这里感觉不用分点写contribution了,因为通常contribution都是我们做了什么，而不是提出的模型表现得怎样。可以直接按照effectiveness，efficency两方面总结在DL和BEIR上的结果，然后写We further evaluate TourRank using a range of large language models (LLMs), including GPT-3.5-turbo, GPT-4 turbo, GPT-4o-mini via OpenAI’s API, as well as several open-source models, such as Mistral-7B-Instruct-v0.2 \cite{jiang2023mistral}, Llama-3-8B-Instruct \cite{Meta2024} and vicuna-13b-v1.5 \cite{zheng2023judging}. The results suggest that TourRank consistently outperform some existing listwise ranking approaches.

% To conclude, our contributions can be summarized as follows:
% % \vspace{-1mm}
% \begin{itemize}%[leftmargin=*]
% % \setlength{\itemsep}{-2pt} % 设置item之间的垂直间距
% % \setlength{\parsep}{-2pt} % 设置段落间的垂直间距
% \item We introduce TourRank, a novel zero-shot documents ranking method based on LLMs, which contains multiple parallel rounds of tournament.
% \item TourRank effectively mitigates the shortcomings of current methods, particularly their sensitivity to the initial candidate documents.
% \item TourRank strikes a commendable balance between inference cost and effectiveness, further solidifying its advantages.
% \item Our experimental results confirm that TourRank not only achieves state-of-the-art on the OpenAI's API, gpt-3.5-turbo, gpt-4-turbo and gpt-4o-mini, but also on some open-source models, such as Mistral-7B-Instruct-v0.2 \cite{jiang2023mistral}, Llama-3-8B-Instruct \cite{Meta2024} and vicuna-13b-v1.5 \cite{zheng2023judging}.
% \end{itemize}

\section{Related Works}

With the development of pre-trained language models like BERT \cite{devlin2018bert} and T5 \cite{raffel2020exploring}, researchers have leverage them in document ranking \cite{fan2022pre}. Notably, \citet{nogueira2019passage} develop a multi-stage text ranking system using BERT, while \citet{nogueira2020document} and \citet{zhuang2023rankt5} employ T5 for document ranking. With the emergence of large language model (LLM), recent studies have utilized LLMs for ranking tasks, employing pointwise, pairwise, and listwise approaches. Pointwise methods, such as Query Generation (QG) \cite{sachan2022improving} and Binary Relevance Generation (B-RG) \cite{liang2022holistic}, use LLMs to compute the probability or likelihood of query-passage pairs. Pairwise approaches, such as Pairwise Ranking Prompting (PRP) \cite{qin2023large}, leverage LLMs to conduct pairwise comparisons and ranking of retrieved documents. \citet{luo2024prp} propose PRP-Graph which utilizes a scoring Pairwise Ranking Prompting unit to construct a ranking graph and aggregates it to enhance LLMs in re-ranking tasks. RankGPT \cite{sun2023chatgpt} is a listwise method that adopts a sliding window strategy for document ranking. Setwise prompting \cite{zhuang2023setwise} enhances efficiency by reducing model inferences and prompt token consumption. ListT5 \cite{yoon2024listt5} is a reranking approach that uses Fusion-in-Decoder architecture and tournament sort for efficiency and effectiveness, and we specifically clarify the differences between ListT5 and our approach in Appendix \ref{LLMs Approaches}.

% There are also other listwise methods, like RankVicuna \cite{pradeep2023rankvicuna} and RankZephyr \cite{pradeep2023rankzephyr}, which employ instruction-tuning for documents ranking.

More introduction of existing works can be seen in Appendix \ref{More Related Works}.

\section{Method: TourRank}

In this section, we introduce a novel zero-shot ranking approach called TourRank, which is inspired by the tournament mechanism. To overcome the limited input length of LLMs and improve the ranking speed, we propose parallel multi-tournaments with multi-stage grouping. And the independent accumulated points system helps TourRank achieve a faster and more robust ranking.

% Similar to how players are ranked based on the accumulated points of multiple tournaments in descending order in a season, TourRank gets the ranking order of candidate documents based on the accumulated points of multi-round tournaments in descending order. 

Next, we first delineate how a basic tournament works in TourRank. Then, we explain how to get the accumulated points of the candidate documents, which are subsequently utilized for document ranking. Lastly, we propose a specific grouping method to circumvent the constraints on the input length of LLMs and make full use of the initial ranking order.

\subsection{A Basic Tournament}

\begin{figure*}[t]
  \centering
  \includegraphics[width=1.0\textwidth]{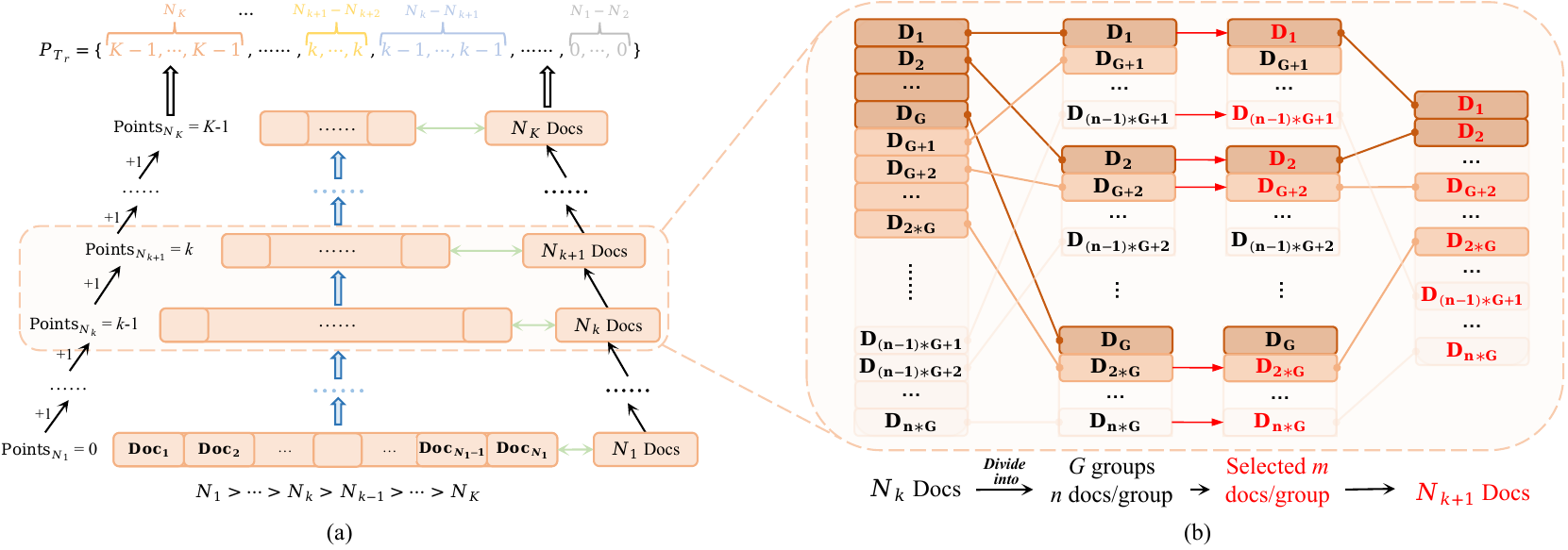}
  \vspace{-8mm}
  \caption{(a) A basic tournament that selects the $N_k$ most relevant documents from $N_1$ candidates with $K$ stages. $P_{T_r}$ is the points vector for all candidates obtained in the $K$ stages. (b) The grouping strategy in the selection stage of the tournament.}
  \vspace{-3.5mm}
  \label{tournament_grouping}
\end{figure*}

For each query, the TourRank approach runs $R$ rounds of tournaments to rank $N_1$ candidate documents retrieved by the first-stage retrieval model. In one tournament of TourRank, we select $N_K$ documents from $N_1$ candidates in a process that consists of $K$ sequential stages and each document gets a corresponding point after a whole tournament. As shown in Figure \ref{tournament_grouping} (a), we choose the documents by stagewise selection ($N_1\rightarrow N_2 \rightarrow \cdots \rightarrow N_{K-1} \rightarrow N_K$). In the $k$-th selection stage ($k \in \{1, 2, \cdots K-1\}$), the top-$N_{k+1}$ documents to the given query are selected from $N_k$ documents to next selection stage. We add 1 point to each document whenever it is selected to advance to the next stage. In this way, after a full round of tournament, all candidate documents can get the corresponding points. As shown in Figure \ref{tournament_grouping} (a) and Table \ref{points},  the $N_k - N_{k+1}$ documents that have qualified for the $k$-th stage but fails to advance will get $k-1$ points. As a special case, the $N_k$ most relevant documents that champion in the $K$-stage tournament will get $K-1$ points. We denote the points vector obtained in the $r$-th round of the tournaments as $P_{T_r}$.

% $P_{T_r}$ which is expressed as Table \ref{points}. In our experiments, the number of candidate documents is 100, and the specific points of all 100 documents after one tournament are shown in Table \ref{specific points} in Appendix \ref{hyperparameters}.

\begin{table}[h] \small %\tiny %\tiny \scriptsize \footnotesize \small \normalsize
\caption{The points of all candidate documents after one tournament. For example, there are $N_{k}-N_{k+1}$ documents with a score of $k-1$. ($k \in \{1, 2, \cdots K-1\}$)}
\centering
\resizebox{0.28\textwidth}{!}{
\begin{tabular}{c|c}
\toprule
\textbf{Number of Docs} & \textbf{Points of Docs} \\
\midrule
$N_K$ & $K-1$ \\
\midrule
$N_{K-1}-N_K$ & $K-2$ \\
\midrule
$\cdots$ & $\cdots$ \\
\midrule
$N_{k}-N_{k+1}$ & $k-1$ \\
\midrule
$\cdots$ & $\cdots$ \\
\midrule
$N_{1}-N_2$ & $0$ \\
\bottomrule
\end{tabular}
}
\label{points}
\end{table}

% The parameter $r$ of $P_{T_r}$ represents the $r$-th round of tournaments. As shown in Figure \ref{multi_filtrations}, $R$ rounds of tournaments can be performed in parallel, so we have $r \in \{ 1, 2, \cdots, R \} $.

\begin{figure}[t]
  \centering
  \includegraphics[width=0.4\textwidth]{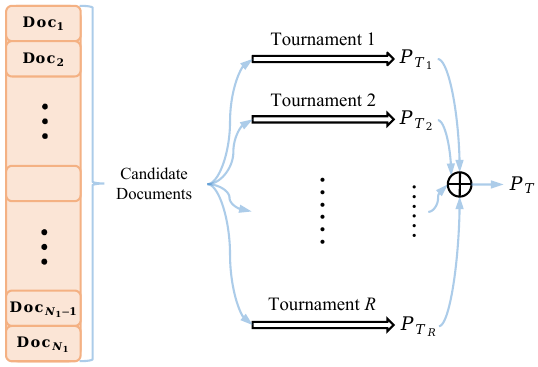}
  \vspace{-2mm}
  \caption{Get the accumulated points of all candidate documents through $R$ tournaments.}
  \label{multi_filtrations}
\end{figure}

\subsection{Getting The Accumulated Points}

Since the points obtained by one tournament are coarse, multiple tournaments are required to obtain more fine-grained document points. Figure \ref{multi_filtrations} illustrates the process of multiple tournaments, where we can see that points of candidate documents $P_{T_r} (r \in \{1, \cdots, R\})$ are obtained after each round of the tournament. 

Because there are many factors that affect the output content of LLMs, such as decoding strategy, temperature coefficient, and especially the order of documents input to LLMs, may introduce some bias, so each set of points vectors ($P_{T_1}, \cdots, P_{T_R}$) obtained by $R$ rounds of tournaments are a little bit different. If these points are added up, the variance of each round tournament could be reduced to some extent, and the accumulated points $P_{T}$, which is expressed as Equation (\ref{pt}), are more fine-grained and robust. So the final ranking list is obtained according to the accumulated points $P_{T}$ in descending order. The analysis in Appendix \ref{Case Study: How Does TourRank Improve the Performance of Documents Ranking?} shows how TourRank-$r$ improves document ranking.

\vspace{-2mm}

\begin{equation}
P_{T} = \sum_{r=1}^{R} P_{T_r}
\label{pt}
\end{equation}

\vspace{-2mm}

\subsection{The Grouping and Selection Strategy}

Considering the limitation of the input length of LLMs, in some stages of TourRank, such as the stage of selecting $N_{k+1}$ documents from $N_{k}$ candidates in Figure \ref{tournament_grouping} (a), we may not be able to input all $N_{k}$ documents into LLMs at once. Therefore, we take the approach of assigning $N_{k}$ candidate documents to several groups and then parallelly prompt LLMs to select top relevant documents within each group, respectively. Such a grouping strategy is similar to the group stage in a sport tournament. 

As shown in Figure \ref{tournament_grouping} (b), the $N_{k}$ documents are divided into $G$ groups, each of which contains $n$ documents. Here the relative order of $N_{k}$ initial documents is given by the retrieval model, such as BM25 \cite{robertson2009probabilistic}, etc. When grouping in a sports tournament, the seeded players and the weaker players are evenly assigned into different groups to ensure the fairness of the competition. Similarly, we used a similar strategy to group the documents by evenly distributing the documents in the initial order into different groups as shown in Figure \ref{tournament_grouping} (b). In this way, there will be some difference in the relevance of the documents within a group, making it easier for LLMs to select the more relevant documents. 

Additionally, \citet{liu2024lost} find that current language models do not robustly access and use information in long input contexts because of the position bias. In order to eliminate the bias of LLMs on document input order and achieve a robust ranking, the order of documents in each group will be shuffled before entering LLMs and the multiple tournaments will be performed as shown in Figure \ref{multi_filtrations}.

After grouping the documents, we select the most relevant $m$ documents from the $n$ ($m<n$) documents in each group. In Figure \ref{tournament_grouping} (b), we mark the selected $m$ documents in red in each group, and these documents advance to the next stage. 

Eventually, through the $k$-th selection stage of the tournament, $N_{k+1}$ more relevant documents are selected from the $N_{k}$ documents to advance to the next stage. Benefiting from this smart grouping stage and multi-round tournaments mechanism, we solve the problem of limited input length of LLMs while achieving a more robust selection. 

\subsection{The Overall of TourRank}

\begin{algorithm} \small %\small %\tiny \scriptsize
\caption{The Pseudo-code of TourRank}
\begin{algorithmic}[1]
% \STATE \textbf{TOURRANK(Query, Docs, LLM)}
\STATE \textbf{Input: The query $q$ and candidate documents list $D$} \vspace{0.5mm}
\STATE \textit{Perform $R$ tournaments \textbf{in parallel}, $r \in \{1, \cdots, R\}$:} \vspace{0.5mm}
\STATE \hspace{\algorithmicindent} \textit{Initialize the points as $\boldsymbol{P_{T_{r}}=0}$ for $N_{1}$ documents.} \vspace{0.5mm}
\STATE \hspace{\algorithmicindent} \textit{Perform $k$-th selection stages, for $k$ in range($1, K$):} \vspace{0.5mm}
\STATE \hspace{\algorithmicindent} \hspace{\algorithmicindent} \textit{Assign $N_{k}$ documents to $G$ groups and each group} \\
\hspace{\algorithmicindent} \hspace{\algorithmicindent} \textit{has $n$ documents.} \vspace{0.5mm}
\STATE \hspace{\algorithmicindent} \hspace{\algorithmicindent} \textit{Select $m$ documents that are more relevant to the} \\
\hspace{\algorithmicindent} \hspace{\algorithmicindent} \textit{query $q$ from $n$ in each group \textbf{in parallel}.} \vspace{0.5mm}
\STATE \hspace{\algorithmicindent} \hspace{\algorithmicindent} \textit{Get the selected $N_{k+1}$ documents to advance to next stage.} \vspace{0.5mm}
\STATE \hspace{\algorithmicindent} \hspace{\algorithmicindent} \textit{The points $\boldsymbol{P_{T_{r}}}$ of the selected $N_{k+1}$ documents add $1$.} \vspace{0.5mm}
\STATE \hspace{\algorithmicindent} \textit{Get a set of points $\boldsymbol{P_{T_{r}}}$ for all $N_{1}$ documents.} \vspace{0.5mm}
\STATE \textit{After $R$ times parallel tournaments, the final points $\boldsymbol{P_{T}}$ can be obtained according to Equation (\ref{pt}).} \vspace{0.5mm}
\STATE \textit{Rank the candidate documents $D$ according to $\boldsymbol{P_{T}}$ in descending order.} \vspace{0.5mm}
\STATE \textbf{Output: A re-ranked list of candidate documents $D_{ranked}$} %\vspace{0.5mm}
\end{algorithmic}
\end{algorithm}

As the Pseudo-code of TourRank shown in \textbf{Algorithm 1}, we perform $R$ parallel tournaments as the process in Figure \ref{multi_filtrations} for the given query $q$ and the candidate documents list $D$. In $r$-th round tournament, we first initialize the points of all $N_1$ candidate documents, that is $P_{T_{r}}=0$ for $N_1$ documents. Then, we select and increase the points of the documents in a stage-by-stage way in which $K-1$ times selection stages are executed serially, and this is corresponds to Figure \ref{tournament_grouping} (a). In $k$-th selection stage, we adopt a suitable grouping approach (Figure \ref{tournament_grouping} (b)) to get the $N_{k+1}$ documents which can advance to the next selection stage, while adding points to the selected $N_{k+1}$ documents. After $R$ rounds tournament, the points $P_{T_{r}}, r \in \{1, \cdots, R\}$ can be obtained. We can calculate the final points $P_{T}$ according to Equation (\ref{pt}). Finally, we re-rank the candidate documents list according to the final points $P_{T}$ in descending order.

The specific hyperparameters of TourRank can be seen in Table \ref{Hyperparameters of TourRank} in the Appendix \ref{hyperparameters}.

\section{Experiments}

Our experiments mainly focus on the following research questions:
% \vspace{-2mm}
\begin{itemize}[leftmargin=*]
\item \textbf{RQ.1}: How does TourRank perform in ranking tasks?
\item \textbf{RQ.2}: How robust is TourRank to the first-stage retrieval models and the initial orders of candidate documents?
% Does TourRank exhibit sensitivity to the candidate documents retrieved by different models and the initial order of documents, essentially, does it maintain a robust ranking?
\item \textbf{RQ.3}: What is the trade-off between ranking effectiveness and computational cost/efficiency when using TourRank for document ranking?
\item \textbf{RQ.4}: How does TourRank perform when using different LLMs, including both open-source models and closed API-based models?
% \item \textbf{RQ.5}: How exactly does TourRank improve the performance of LLMs on documents ranking?
\end{itemize}

\subsection{Experimental Settings}

\subsubsection{Datasets}

We conduct experiments to answer the above research questions on TREC DL datasets \cite{craswell2020overview, craswell2021overview} and BEIR benchmark \cite{thakur2021beir}. \textbf{TREC} is a widely used benchmark in IR research. We use the test sets of TREC DL 19 and TREC DL 20, which contain 43 and 54 queries. \textbf{BEIR} is a heterogeneous zero-shot evaluation benchmark. Following~\citet{sun2023chatgpt}, we select 8 datasets for evaluation, including Covid, Touche, DBPedia, SciFact, Signal, News, Robust04, and NFCorpus.

\subsubsection{Metrics}

In the next evaluations, we re-rank the top-100 documents retrieved by the first-stage retrieval model. If not specified, we use BM25 as the default retrieval model and PySerini for implementation.\footnote{https://github.com/castorini/pyserini} We use NDCG@\{5, 10, 20\} as evaluation metrics.

\subsubsection{Baselines}

We compare TourRank with several baselines in documents ranking, including the supervised methods:

\begin{itemize}[leftmargin=*]
\item \textbf{monoBERT} \cite{nogueira2019passage}: A ranking method with a cross-encoder architecture based on BERT-large, trained on MS MARCO.
\item \textbf{monoT5} \cite{nogueira2020document}: A ranking method that calculates the scores using T5 model.
% \item \textbf{TART} \cite{asai2022task}: TART advances zero-shot retrieval by tuning to multi-task instructions from the BERRI dataset.
\end{itemize}

And the zero-shot methods based on LLMs:

\begin{itemize}[leftmargin=*]
\item \textbf{DIRECT(0, 10)} \cite{guo2024generating}: A pointwise method which gives the relevance scores ranging from 0 to 10 to each query-document pair in text format using LLMs. Then, rank the documents according to these scores in descending order.
% \item \textbf{Query Generation (QG)} \cite{sachan2022improving}: A pointwise method which uses LLM to compute the probability of the given query conditioned on a retrieved document.
\item \textbf{Binary Relevance Generation (B-RG)} \cite{liang2022holistic}: A pointwise method which ranks the candidate documents according to the likelihood of "Yes or No" on a query-document pair.
\item \textbf{PRP} \cite{qin2023large}: A pairwise method that reduces the burden on LLMs by using a technique called Pairwise Ranking Prompting.
\item \textbf{Setwise} \cite{zhuang2023setwise}: A listwise method that improves the efficiency of LLM-based zero-shot ranking. The authors introduce two Setwise methods, Setwise.bubblesort and Setwise.heapsort. We reproduce the performance of Setwise.heapsort and Setwise.bubblesort based on the code publicly available on Github \footnote{https://github.com/ielab/llm-rankers} in the original Setwise paper. The Setwise paper mentioned that the hyperparameter $c=10$ is the best value for gpt-3.5-turbo API, so we also use $c=10$ in our experiments. Here $c$ refers to the number of documents compared in a prompt.
\item \textbf{RankGPT} \cite{sun2023chatgpt}: A listwise method that uses a sliding window strategy to achieve listwise ranking based on LLMs. The experiments are also based on the code publicly available in the original paper \footnote{https://github.com/sunnweiwei/RankGPT}.
\end{itemize}

\definecolor{color0}{rgb}{0.93,0.93,0.93} % 最浅
\definecolor{color1}{rgb}{0.895,0.895,0.895}
\definecolor{color2}{rgb}{0.85,0.85,0.85}
\definecolor{color3}{rgb}{0.755,0.755,0.755} % 最深
\definecolor{color4}{rgb}{0.615,0.615,0.615} % 最深

\begin{table*}[t!] \footnotesize %\footnotesize % \footnotesize
\caption{Performance comparison of different methods on TREC datasets. We reproduce all the zero-shot LLM methods with gpt-3.5-turbo API. The best-performing algorithms for supervised methods and zero-shot LLM methods are bolded, respectively. The best top-4 results of zero-shot LLM methods are shaded in each metric. TourRank-$r$ represents that we perform $r$ times tournaments.}
% \vspace{-3mm}
\centering
\resizebox{0.79\textwidth}{!}{%
\begin{tabular}{@{}c|ccc|ccc@{}}
\toprule
\multirow{2}{*}{\textbf{Methods}} & \multicolumn{3}{c|}{\textbf{TREC DL 19}} & \multicolumn{3}{c}{\textbf{TREC DL 20}} \\
 & \textbf{NDCG@5} & \textbf{NDCG@10} & \textbf{NDCG@20} & \textbf{NDCG@5} & \textbf{NDCG@10} & \textbf{NDCG@20} \\
\midrule
BM25 & 52.78 & 50.58 & 49.14 & 50.67 & 47.96 & 47.21 \\
\midrule
\multicolumn{7}{c}{\textbf{Supervised Methods}} \\
\midrule
monoBERT (340M) & 73.25 & 70.50 & - & 70.74 & 67.28 & - \\
monoT5 (220M) & \textbf{73.77} & 71.48 & - & 69.40 & 66.99 & - \\
monoT5 (3B) & 73.74 & \textbf{71.83} & - & \textbf{72.32} & \textbf{68.89} & - \\
% \addlinespace
\midrule
\multicolumn{7}{c}{\textbf{Zero-Shot LLM Methods}} \\
\midrule
DIRECT(0, 10) & 54.22 & 54.59 & 54.15 & 55.17 & 55.35 & 54.73 \\
% QG & gpt-3.5-turbo & 69.06 & 65.93 & 60.82 & 66.72 & 63.71 & 58.87 \\
B-RG & 63.33 & 62.51 & 60.00 & 65.04 & 63.37 & 60.47 \\
PRP-Allpair & 70.43 & \cellcolor{color0}68.18 & \cellcolor{color0}64.61 & \cellcolor{color3}69.75 & \cellcolor{color3}66.40 & \cellcolor{color3}64.03 \\
Setwise.heapsort (c=10) & 70.55 & 68.16 & \cellcolor{color3}65.63 & 57.05 & 53.73 & 51.66 \\
Setwise.bubblesort (c=10) & 67.62 & 66.19 & 63.41 & 57.03 & 53.82 & 50.79 \\
RankGPT & \cellcolor{color2}72.05 & \cellcolor{color2}68.19 & 62.21 & \cellcolor{color0}67.25 & 63.60 & 59.12 \\
% RankGPT & \cellcolor{color4}75.98 & \cellcolor{color3}72.67 & \cellcolor{color1}67.31 & \cellcolor{color2}72.32 & \cellcolor{color2}69.48 & \cellcolor{color1}63.53 \\
\midrule
% \multicolumn{8}{c}{\textbf{TourRank (Ours)}} \\
% \midrule
TourRank-1 & \cellcolor{color0}70.95 & 66.23 & 62.49 & 66.65 & \cellcolor{color0}63.74 & \cellcolor{color0}60.59 \\
TourRank-2 & \cellcolor{color3}72.24 & \cellcolor{color3}69.54 & \cellcolor{color2}65.03 & \cellcolor{color2}67.65 & \cellcolor{color2}65.20 & \cellcolor{color2}62.78 \\
TourRank-10 & \cellcolor{color4}\textbf{73.83} & \cellcolor{color4}\textbf{71.63} & \cellcolor{color4}\textbf{68.37} & \cellcolor{color4}\textbf{72.49} & \cellcolor{color4}\textbf{69.56} & \cellcolor{color4}\textbf{66.13} \\
% TourRank-1 & gpt-4-turbo & \cellcolor{color2}73.99 & \cellcolor{color2}72.46 & \cellcolor{color2}67.63 & \cellcolor{color1}70.69 & 67.38 & \cellcolor{color2}63.94 \\
% TourRank-5 & gpt-4-turbo & \cellcolor{color3}75.57 & \cellcolor{color4}74.13 & \cellcolor{color4}70.04 & \cellcolor{color3}72.46 & \cellcolor{color3}69.79 & \cellcolor{color4}65.61 \\
\bottomrule
\end{tabular}
}
\label{TREC}
\end{table*}

\subsection{Experimental Results}

\textbf{Results on TREC DL datasets} \quad  Table \ref{TREC} shows the performance of different methods on TREC DL datasets. We compare NDCG@\{5, 10, 20\}, and the best top-4 results of zero-shot LLM methods are shaded. We reproduce all zero-shot LLM methods with gpt-3.5-turbo API. 
From the results, we can make the following findings: 

\noindent
\textbf{(1)} Our TourRank-10 outperforms all zero-shot ranking baselines. It is worth noting that after two tournaments (TourRank-2) the performance is much better than one tournament (TourRank-1), and TourRank-2 can significantly outperform RankGPT and Setwise. This indicates that TourRank can achieve good results with fewer tournaments. 
% In addition, although the pointwise approach, PointWise and B-RG, consume fewer tokens, it comes at the cost of poor results; 

\noindent
\textbf{(2)} Generally, the two pointwise methods, DIRECT(0, 1) and B-RG, tend to underperform in comparison to the pairwise and listwise methods. This is because the pointwise methods evaluate each document independently to determine whether it is relevant to the query, while pairwise and listwise methods compare each document to several other documents simultaneously.

\noindent
\textbf{(3)} PRP-Allpair achieves about the same performance as Setwise.heapsort (c=10) and RankGPT on TREC DL 19, and outperforms RankGPT on TREC DL 20. However, PRP-Allpair achieve relatively good results at the cost of much higher complexity and resource consumption than RankGPT and TourRank. We discuss the effectiveness and cost of them in Section~\ref{The Trade-Off between Effectiveness and Resource Consumption}.

\noindent
\textbf{(4)} TourRank-10 achieves comparable results to the best supervised methods on TREC DL 19, and on TREC DL 20 TourRank-10 outperforms the best performing supervised method monoT5 (3B). It can be seen that TourRank is the only zero-shot method based on gpt-3.5-turbo API that can do this. We also perform TourRank-$r$ with other close and open-source LLMs in Section \ref{TourRank Based on Other LLMs}.

% \textbf{(ii)} In methods based on gpt-4-turbo, TourRank-5 outperforms RankGPT in all other metrics except NDCG@5 on TREC DL 19. And TourRank-5 is more than 2\% higher than RankGPT on NDCG@20, even TourRank-1 has higher NDCG@20 than RankGPT. This shows that RankGPT is better at ranking the top-10 documents and does not pay much attention to ranking the later documents. This is also determined by the sliding window strategy of RankGPT. While our TourRank uses the tournament mechanism and thus improves the overall document ranking, not just the top-k documents. The comparison of more metrics can be seen in the Appendix \ref{more_metrics}; 
% \textbf{(iii)} PRP-Allpair is a pairwise method based on FLAN-T5-XXL and has good performance on both datasets, and especially outperforms all methods based on gpt-4-turbo on the DL20 dataset. However, PRP works well at the cost of consuming a particularly large number of tokens. We talk about the cost of different methods in the next Section \ref{The Trade-Off between Effectiveness and Resource Consumption}.

\begin{table*}[t!] \small
\caption{Performance (NDCG@10) comparison of different methods on BEIR benchmark. The best-performing algorithms for supervised methods and zero-shot LLM methods are bolded. TourRank-$r$ represents that we perform $r$ times tournaments.}
% \vspace{-3mm}
\centering
\resizebox{0.84\textwidth}{!}{%
\begin{tabular}{c|cccccccc|c}
\toprule
\textbf{Methods} & \textbf{Covid} & \textbf{NFCorpus} & \textbf{Touche} & \textbf{DBPedia} & \textbf{SciFact} & \textbf{Signal} & \textbf{News} & \textbf{Robust04} & \textbf{Average}\\
\midrule
BM25 & 59.47 & 30.75 & \textbf{44.22} & 31.80 & 67.89 & \textbf{33.05} & 39.52 & 40.70 & 43.42 \\
\midrule
\multicolumn{10}{c}{\textbf{Supervised Methods}} \\
\midrule
monoBERT (340M) & 70.01 & 36.88 & 31.75 & 41.87 & 71.36 & 31.44 & 44.62 & 49.35 & 47.16 \\
monoT5 (220M) & 78.34 & 37.38 & 30.82 & 42.42 & 73.40 & 31.67 & 46.83 & 51.72 & 49.07 \\
monoT5 (3B) & \textbf{80.71} & \textbf{38.97} & 32.41 & \textbf{44.45} & \textbf{76.57} & 32.55 & \textbf{48.49} & \textbf{56.71} & \textbf{51.36} \\
% TART & FLAN-T5-XL (3B) & 75.10 & 36.03 & 27.46 & 42.53 & 74.84 & 25.84 & 40.01 & 50.75 & 46.57 \\
% \addlinespace
\midrule
\multicolumn{10}{c}{\textbf{Zero-Shot LLM Methods}} \\
\midrule
% PointWise & FLAN PaLM2 S & \cellcolor{color0}79.30 & \cellcolor{color2}37.52 & 25.22 & 40.82 & 70.08 & 29.12 & 46.19 & 53.78 & 47.75 \\
% B-RG & FLAN PaLM2 S & 78.97 & \cellcolor{color1}37.43 & 24.27 & 36.96 & 69.58 & \cellcolor{color0}31.96 & 45.88 & \cellcolor{color1}56.56 & 47.70 \\
% PRP-Allpair & FLAN-T5-XXL (11B) & \cellcolor{color1}79.62 & - & 29.81 & 41.41 & \cellcolor{color2}74.23 & \cellcolor{color2}32.22 & 47.68 & \cellcolor{color2}56.76 & - \\
% Setwise.bubblesort & FLAN-T5-XXL (11B) & 76.80 & 34.60 & \cellcolor{color4}38.80 & \cellcolor{color1}42.40 & \cellcolor{color4}75.40 & \cellcolor{color4}34.30 & \cellcolor{color0}47.90 & 53.40 & \cellcolor{color2}50.50 \\

RankGPT & 76.67 & 35.62 & 36.18 & 44.47 & 70.43 & 32.12 & 48.85 & 50.62 & 49.37 \\
\midrule
TourRank-1 & 77.17 & 36.35 & 29.38 & 40.62 & 69.27 & 29.79 & 46.41 & 52.70 & 47.71 \\
TourRank-2 & 79.85 & 36.95 & 30.58 & 41.95 & 71.91 & 31.02 & 48.13 & 55.27 & 49.46 \\
TourRank-10 & \textbf{82.59} & \textbf{37.99} & 29.98 & \textbf{44.64} & \textbf{72.17} & 30.83 & \textbf{51.46} & \textbf{57.87} & \textbf{50.94} \\

% RankGPT & 76.67 & 35.62 & \cellcolor{color2}36.18 & \cellcolor{color2}44.47 & 70.43 & \cellcolor{color1}32.12 & \cellcolor{color2}48.85 & 50.62 & \cellcolor{color0}49.37 \\
% \midrule
% TourRank-1 & 77.17 & 36.35 & 29.38 & 40.62 & 69.27 & 29.79 & 46.41 & 52.70 & 47.71 \\
% TourRank-2 & \cellcolor{color2}79.85 & \cellcolor{color0}36.95 & \cellcolor{color1}30.58 & \cellcolor{color0}41.95 & \cellcolor{color0}71.91 & 31.02 & \cellcolor{color1}48.13 & \cellcolor{color0}55.27 & \cellcolor{color1}49.46 \\
% TourRank-10 & \cellcolor{color4}82.59 & \cellcolor{color4}37.99 & \cellcolor{color0}29.98 & \cellcolor{color4}44.64 & \cellcolor{color1}72.17 & 30.83 & \cellcolor{color4}51.46 & \cellcolor{color4}57.87 & \cellcolor{color4}50.94 \\
\bottomrule
\end{tabular}
}
\label{BEIR}
\end{table*}

% \subsubsection{Results on BEIR benchmark}

\noindent
\textbf{Results on BEIR benchmark} \quad Table \ref{BEIR} shows the NDCG@10 of different methods on 8 tasks of BEIR benchmark. Due to the cost limitation, we compare two zero-shot LLM methods, RankGPT and TourRank-$r$, on BEIR benchmark. The following are some valuable discussions: 

\noindent
\textbf{(1)} TourRank-10 achieves the best performance in 6 out of 8 tasks and the best average NDCG@10 across 8 tasks among zero-shot LLM methods.

\noindent
\textbf{(2)} The average of TourRank-2 (49.46) outperforms RankGPT (49.37) in terms of NDCG@10 in Table~\ref{BEIR}, which, together with the better performance of TourRank-2 over RankGPT on TREC DL datasets shown in Table \ref{TREC}, prove that our TourRank algorithm can achieve good results with only a few times tournaments.

\noindent
\textbf{(3)} Note that on the Touche task and Signal task, all supervised methods and zero-shot methods in Table~\ref{BEIR} perform even worse than BM25. The NDCG@10 of all methods on these two tasks is low, only about 0.3. According to \citet{thakur2024systematic}, the poor performance of neural retrieval models is mainly due to the large number of short texts and unlabeled texts in the Touche dataset.

The results on TREC datasets and BEIR benchmark jointly answer the \textbf{RQ.1}.

\subsection{Sensitivity Analysis to Initial Ranking}
\label{Sensitivity Analysis to Initial Ranking}

\begin{figure}[h]
\centering
\begin{subfigure}{0.5\textwidth}
  \centering
  \includegraphics[width=0.85\linewidth]{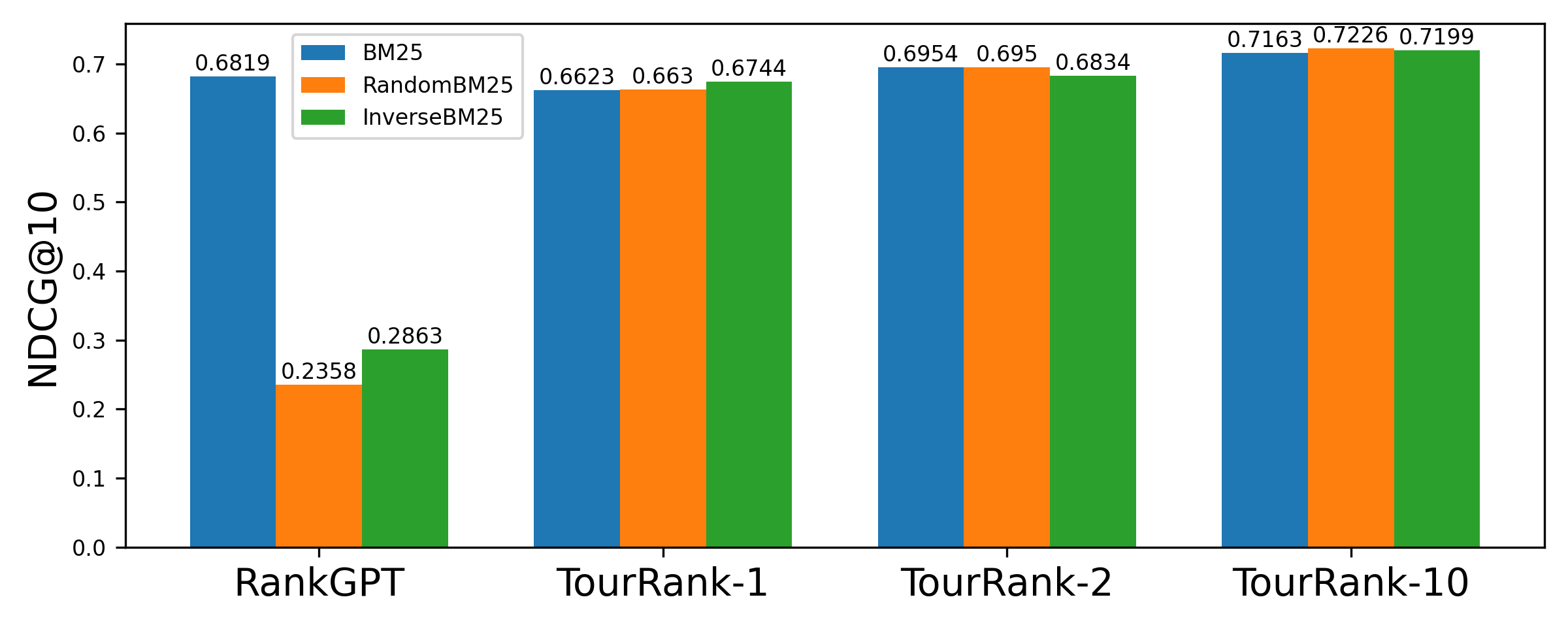}
  \caption{TREC DL 19}
  % \vspace{-1mm}
  \label{sensitivity_dl19}
\end{subfigure}
% \vspace{1cm} % 这里1cm是你想要的距离，你可以更改这个数值
\begin{subfigure}{0.5\textwidth}
  \centering
  \includegraphics[width=0.85\linewidth]{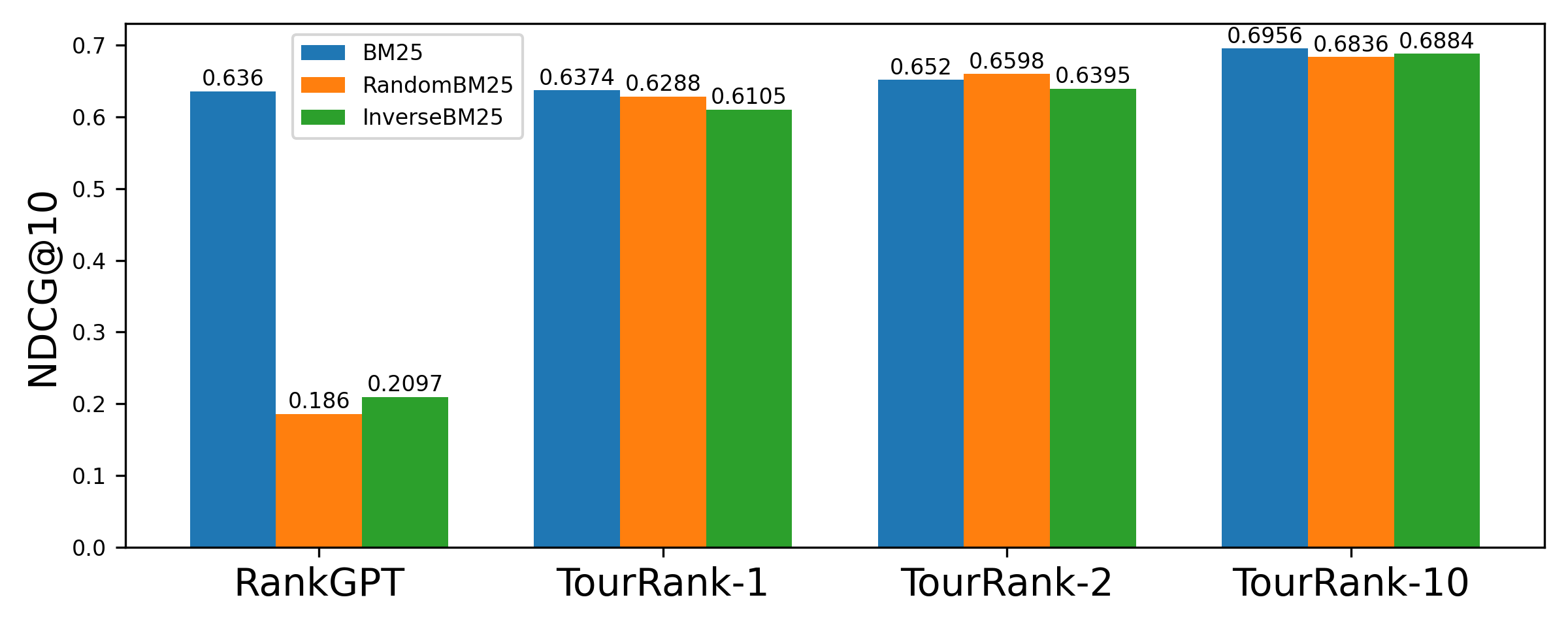}
  % \vspace{-1mm}
  \caption{TREC DL 20}
  \label{sensitivity_dl20}
\end{subfigure}
\vspace{-7mm}
\caption{The sensitivity analysis to initial ranking of TourRank and RankGPT on TREC DL 19 and TREC DL 20.}
\label{sensitivity}
\end{figure}

We compare 3 different initial ranking: 1) \textbf{BM25}: Get top-100 documents by BM25; 2) \textbf{RandomBM25}: Shuffle the order of BM25; 3) \textbf{InverseBM25}: Reverse the order of BM25. Figure \ref{sensitivity} shows the results of RankGPT and our TourRank based on 3 initial rankings and all these experiments are based on gpt-3.5-turbo API. 

From Figure \ref{sensitivity}, we can see that RankGPT is very sensitive to the initial permutation of documents list. When the initial permutation is shuffled or reversed, the performance of RankGPT becomes much worse. This is caused by the ranking mechanism of RankGPT, which adjusts the overall permutation of documents list through the sliding window strategy. Sliding the window from bottom to top makes it easier for documents that are originally near the top to be ranked at top positions in the final permutation. Whereas documents that are at the bottom of the initial permutation need to be ranked at the top of every comparison in corresponding sliding window in order to be ranked at the top of the final permutation, otherwise they are left at the bottom or middle of the whole documents list. So, this is the reason why RankGPT is very sensitive to the initial ranking.

% When TourRank-1 and TourRank-2 are faced with RandomBM25 and InverseBM25 as the initial ranking of candidate documents, the performances are slightly lower than the initial ranking based on BM25. When after 10 tournaments, TourRank-10 achieves almost the same ranking effect based on 3 different initial rankings.

However, our TourRank is quite robust to different initial orderings, as shown by the fact that shuffling and reversing the initial order has almost no effect on TourRank-$r$. The robustness of TourRank to the initial ranking benefits from the tournament mechanism presented in Figure \ref{tournament_grouping}. Each tournament is a selection over all candidate documents, not just a fine-tuning of the initial ranking like RankGPT. 

% Moreover, the ranking based on the accumulated points $P_{T}$ of multiple parallel tournaments (Figure \ref{multi_filtrations}) can make the ranking results given by TourRank more robust. 

% These analyses to the sensitivity of TourRank answer the \textbf{RQ.2}.

\subsection{Analysis to Different Retrieval Models}

In addition to BM25, we also obtain top-100 documents based on two more powerful retrieval models, including a dense retriever model Contriever \cite{izacard2021unsupervised} and a neural sparse retrieval model SPLADE++ ED \cite{formal2022distillation}, as the first-stage retrieval model. Then, we perform TourRank and RankGPT to re-rank the top-100 candidate documents retrieved by different retrieval models based on gpt-3.5-turbo API. The results in Table~\ref{initial_ranking} show that TourRank-10 achieves SOTA ranking performance based on 3 kinds of different top-100 initial candidate documents. And TourRank-2 can also outperform RankGPT in general.

\begin{table}[th] %\tiny
\caption{NDCG@10 of TourRank and RankGPT based on different retrieval models. Here we use gpt-3.5-turbo API for TourRank and RankGPT.}
\centering
\resizebox{0.48\textwidth}{!}{%
\begin{tabular}{c|c|cc}
\toprule
\textbf{Methods} & \textbf{Top-100} & \textbf{TREC DL 19} & \textbf{TREC DL 20}\\
\midrule
% \midrule
BM25 & - & 50.58 & 47.96 \\
RankGPT & \multirow{3}{*}{BM25} & 68.19 & 63.60 \\
TourRank-2 & & 69.54 & 65.20 \\
TourRank-10 & & \textbf{71.63} & \textbf{69.56} \\
\midrule
% \midrule
Contriever & - & 62.02 & 63.42 \\
RankGPT & \multirow{3}{*}{Contriever} & 69.70 & 68.47 \\
TourRank-2 &  & 69.12 & 71.89 \\
TourRank-10 &  & \textbf{70.77} & \textbf{73.19} \\
\midrule
% \midrule
SPLADE++ ED & - & 73.08 & 71.97 \\
RankGPT & \multirow{3}{*}{SPLADE++ ED} & 74.56 & 70.75 \\
TourRank-2 &  & 74.86 & 74.11 \\
TourRank-10 &  & \textbf{75.35} & \textbf{77.09} \\
% \midrule
\bottomrule
\end{tabular}
}
\label{initial_ranking}
\end{table}

The results in Table \ref{initial_ranking} and the analysis in Section~\ref{Sensitivity Analysis to Initial Ranking} jointly answer the \textbf{RQ.2}, that is, TourRank has the ability of robust ranking.

\subsection{The Trade-Off between Effectiveness and Resource Consumption}
\label{The Trade-Off between Effectiveness and Resource Consumption}

In order to prove that our TourRank can achieve a good balance between effectiveness and efficiency, we conduct theoretical analysis and comparison of actual cost and latency.

\subsubsection{Theoretical analysis of efficiency}

Table \ref{time_token} shows the approximation of the theoretical lowest time complexity of different methods and the number of documents LLMs need to receive. All the contents of Table \ref{time_token} are based on the recommended parameters. More detailed discussions on precise time complexity and number of input documents are in Table \ref{time_token_2} in the Appendix \ref{The Discussions on Time Complexity and Number of Documents Inputted to LLMs}. From Table \ref{time_token}, we can see that: 

\begin{table}[th] %\normalsize
\caption{A approximation of the theoretical lowest time complexity of various methods and the number of documents which are inputted to LLMs for each method. $N$ is the number of candidate documents. Setwise ranks the top-$k$ ($k<N$) documents through bubblesort and heapsort, and $c=10$ is the documents compared in a prompt of Setwise based on gpt-3.5-turbo API. $K-1$ is the times of the selection stages in a tournament (Figure \ref{tournament_grouping} (a)) and $r$ is the times of tournaments in TourRank-$r$. (Note: The approximate contents in this table are based on the recommended parameters.)}
\centering
\resizebox{0.48\textwidth}{!}{%
\begin{tabular}{ccc}
\toprule
% \hline
\textbf{Methods} & \textbf{Time Complexity} & \textbf{No. of Docs to LLMs} \\ 
\midrule
PointWise & $O(1)$ & $N$ \\ %\hline
\midrule
PRP-Allpair & $O(1)$ & $N^2-N$ \\ %\hline
\midrule
Setwise.bubblesort & $\approx O(\frac{1}{9}k*N)$ & $\approx \frac{10}{9}k*N$ \\ %\hline
\midrule
Setwise.heapsort & $\approx O(k * log_{10} N)$ & $\approx 10 k * log_{10} N$ \\ %\hline
\midrule
RankGPT & $\approx O(\frac{1}{10}*N)$ & $\approx 2*N$ \\ %\hline
\midrule
TourRank-$r$ & $O(K-1)$ & $\approx 2r*N$ \\ 
\bottomrule
\end{tabular}%
}
\label{time_token}
\end{table}

\noindent
\textbf{(1)} Pointwise has the lowest time complexity and the lowest number of documents received by LLMs, but the experimental results of DIRECT(0, 10) and B-RG in Table \ref{TREC} show that PointWise exhibits poor performance. 

\noindent
\textbf{(2)} Although the pairwise method, PRP-Allpair, performs well in the experiments on TREC datasets, the number of input documents required by PRP-AllPair is $N^2-N$, which will greatly increase the cost of ranking. And the theoretical optimal time complexity of pairwise method is $O(1)$ when all pairwise documents can be compared at the same time. However, for example, in order to sort 100 candidate documents and achieve the theoretical optimal $O(1)$ time complexity, $\frac{100 \times 99}{2}=4950$ document pairs should be compared at the same time. While API request rates are often limited in reality, so the actual latency of pairwise method is much higher. We will discuss the actual latency in Section \ref{Cost and Latency in Practice}.

\begin{figure*}[th]
\centering
\begin{subfigure}{0.498\textwidth}
  \centering
  \includegraphics[width=1.00\linewidth]{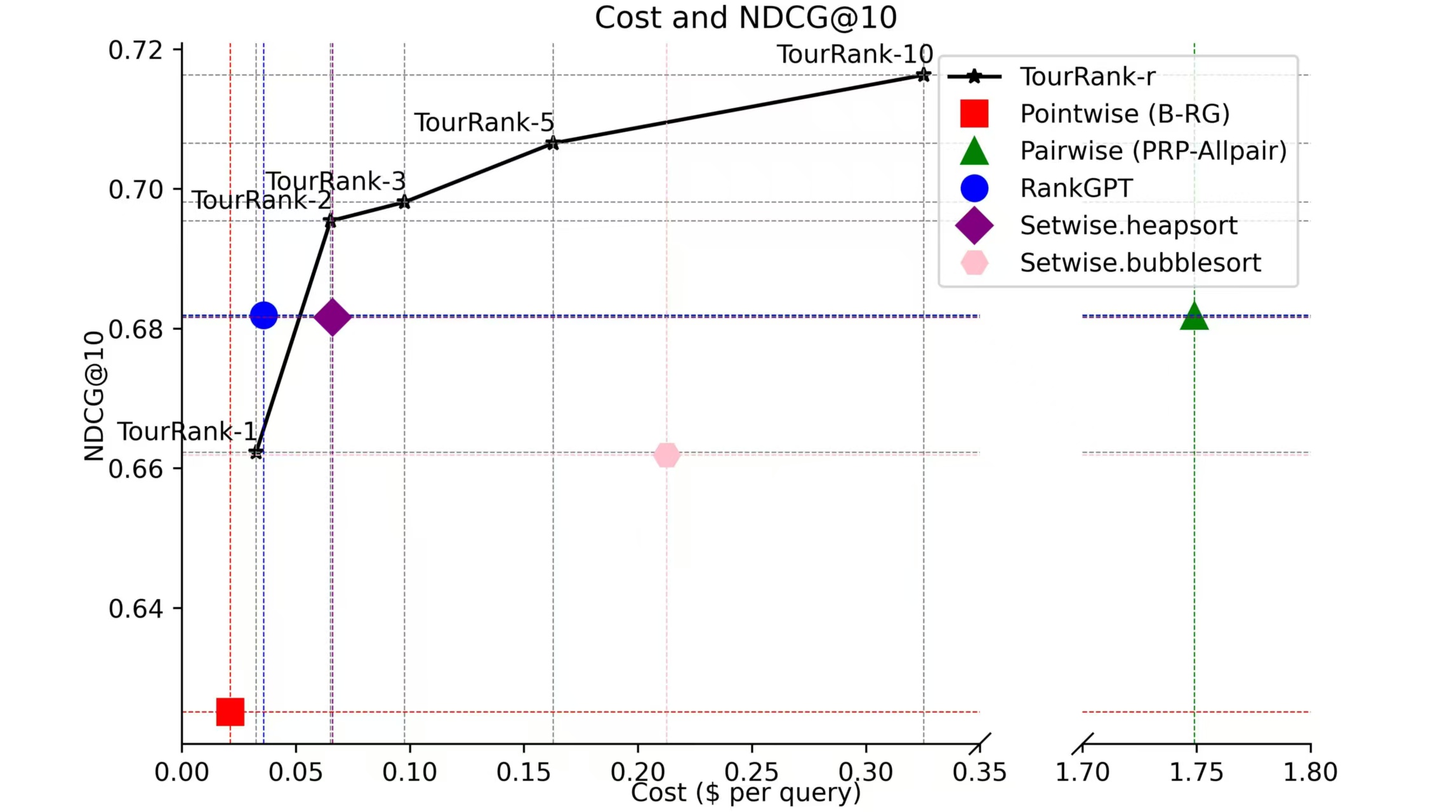}
  \vspace{-0.5mm}
  \caption{Cost vs. NDCG@10}
  \label{Cost vs NDCG@10}
\end{subfigure}
% \vspace{1cm} % 
\begin{subfigure}{0.498\textwidth}
  \centering
  \includegraphics[width=1.00\linewidth]{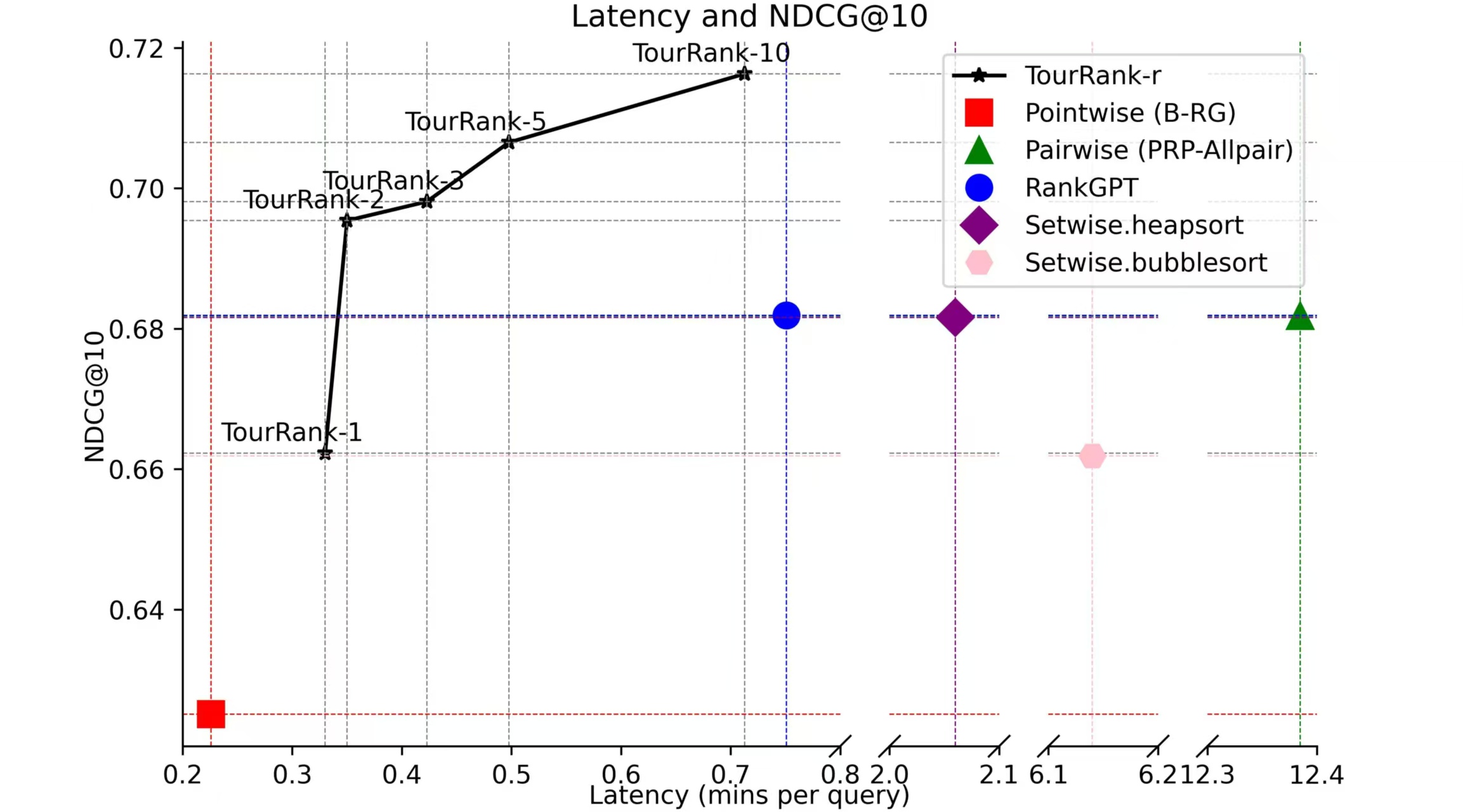}
  \vspace{-0.5mm}
  \caption{Latency vs. NDCG@10}
  \label{Latency vs NDCG@10}
\end{subfigure}
% \vspace{-1.5mm}
\caption{Relationship between Cost / Latency and NDCG@10 on TREC DL 19.}
% \vspace{-1.5mm}
\label{trade_off_cost_latency}
\end{figure*}

\noindent
\textbf{(3)} Setwise performs relatively well on TREC DL 19 dataset in Table \ref{TREC} and can finish the ranking process with inputting fewer documents to LLM comparing to pairwise method. However, the multiple steps of Setwise have dependencies and cannot be run in parallel, resulting in the high time complexity and resource consumption. And We discuss the actual cost and latency in Section \ref{Cost and Latency in Practice}.

\noindent
\textbf{(4)} Two listwise methods RankGPT and our TourRank take into account both the time complexity and the number of documents inputted to LLMs. The experimental results in Table \ref{TREC} and \ref{BEIR} show that TourRank-2 can outperform RankGPT. From Table \ref{time_token}, we can see that TourRank-2 ($r=2$) achieves this goal with about twice as many documents to LLMs as RankGPT but with lower time complexity. We also compare TourRank with running RankGPT multiple iterations in serial (Appendix \ref{Comparison Between Serial RankGPT and Parallel TourRank-$r$}), and TourRank demonstrates better performance and lower consumption.

\subsubsection{Empirical results for ranking cost and latency}
\label{Cost and Latency in Practice}

We compare the actual cost and latency per query of our TourRank-$r$ and several baselines on TREC DL 19 dataset at an API request rate of 30 times per second using gpt-3.5-turbo. Figure \ref{trade_off_cost_latency} provides an intuitive illustration of the relationship between cost, latency, and performance across various methods:

\noindent
\textbf{(1)} The pointwise method B-RG has the lowest actual cost and latency, but NDCG@10 of B-RG is far lower than other methods. It shows that the pointwise method consumes less resources, but it comes at the cost of the worst performance.

\noindent
\textbf{(2)} Although PRP-Allpair perform well in TREC DL datasets in Table \ref{TREC}, the cost and the latency is extremely high comparing to all the other method in Figure \ref{trade_off_cost_latency}. In other words, the pairwise method achieves good results with higher resource consumption, which is relatively low cost performance.

\noindent
\textbf{(3)} In addition, we can see that RankGPT and Setwise.heapsort have similar performance as PRP-Allpair, but the cost and latency are much lower than PRP-Allpair. Moreover, the latency of Setwise.heapsort is nearly three times that of RankGPT, which indicates that RankGPT does a better balance between resource consumption and performance than Setwise.heapsort, and Setwise.heapsort is much better than PRP-Allpair.

\noindent
\textbf{(4)} For our TourRank-$r$, when $r=2$, the NDCG@10 of TourRank-2 has surpassed other baselines, and the cost of TourRank-2 is similar to RankGPT and Setwise.heapsort. The latency of TourRank-$r$ is lower than all baselines except pointwise method, and when $r=10$, the NDCG@10 of TourRank-10 is significantly higher than all baselines. That is, TourRank-$r$ can achieve good performance at low cost and especially lower latency.

Therefore, TourRank-$r$ can achieve very good performance with low resource consumption, which means TourRank can achieve a better trade-off between effectiveness and efficiency in practice.

The above theoretical analysis and the experimental results of actual cost and latency jointly answer the \textbf{RQ.3}.

% \begin{figure}[h]
% \centering
% \begin{subfigure}{0.5\textwidth}
%   \centering
%   \includegraphics[width=0.85\linewidth]{fig/cost.jpg}
%   \caption{Cost vs. NDCG@10}
%   \label{Cost vs NDCG@10}
% \end{subfigure}
% \begin{subfigure}{0.5\textwidth}
%   \centering
%   \includegraphics[width=0.85\linewidth]{fig/latency.jpg}
%   \caption{Latency vs. NDCG@10}
%   \label{Latency vs NDCG@10}
% \end{subfigure}
% \vspace{-7mm}
% \caption{Relationship between Cost / Latency and NDCG@10 on TREC DL 19.}
% \label{trade_off_cost_latency}
% \end{figure}

% \vspace{-5mm}

\subsection{TourRank Based on Other LLMs}
\label{TourRank Based on Other LLMs}

We also explore the effect of zero-shot listwise ranking methods based on different LLMs.  Specifically, we perform TourRank, RankGPT and Setwise based on three open-source models Mistral-7B-Instruct-v0.2 \cite{jiang2023mistral}, Llama-3-8B-Instruct \cite{Meta2024}, vicuna-13b-v1.5 \cite{zheng2023judging}, and OpenAI's API, gpt-4-turbo and gpt-4o-mini. 

% \begin{table}[h] %\tiny
% \centering
% \resizebox{0.48\textwidth}{!}{%
% \begin{tabular}{c|c|cc}
% \toprule
% \textbf{Methods} & \textbf{LLMs} & \textbf{TREC DL 19} & \textbf{TREC DL 20}\\
% \midrule
% % \midrule
% BM25 & - & 50.58 & 47.96 \\
% \midrule
% % \midrule
% % BM25 & \multirow{5}{*}{Llama 3 8B} & 50.58 & 47.96 \\
% RankGPT & \multirow{5}{*}{Llama-3-8B-Instruct} & 59.48 & 54.47 \\
% % TourRank-1 & & 64.92 & 62.22 \\
% Setwise.heapsort & & 58.31 & 44.92 \\
% Setwise.bubblesort & & 49.85 & 34.42 \\
% TourRank-2 & & 71.17 & 66.84 \\
% TourRank-10 & & \textbf{73.30} & \textbf{67.25} \\
% \midrule
% RankGPT & \multirow{5}{*}{vicuna-13b-v1.5} & 63.90 & \textbf{60.80} \\
% % TourRank-1 & & 64.92 & 62.22 \\
% Setwise.heapsort & & 65.90 & 58.30 \\
% Setwise.bubblesort & & 62.20 & 60.20 \\
% TourRank-2 & & 58.55 & 49.37 \\
% TourRank-10 & & \textbf{66.21} & 59.60 \\
% \midrule
% RankGPT & \multirow{3}{*}{Mistral-7B-Instruct-v0.2} & 61.90 & 58.54 \\
% % TourRank-1 & & 62.38 & 57.86 \\
% TourRank-2 & & 65.85 & 62.31 \\
% TourRank-10 & & \textbf{68.64} & \textbf{65.04} \\
% \midrule
% RankGPT & \multirow{3}{*}{gpt-4-turbo} & 72.67 & 69.48 \\
% TourRank-1 & & 72.46 & 67.38 \\
% TourRank-5 & & \textbf{74.13} & \textbf{69.79} \\
% \midrule
% RankGPT & \multirow{3}{*}{gpt-4o-mini} & 72.85 & \textbf{70.35} \\
% TourRank-1 & & 73.35 & 67.89 \\
% TourRank-5 & & \textbf{75.57} & 70.07 \\
% \bottomrule
% \end{tabular}
% }
% \caption{NDCG@10 of TourRank and RankGPT based on open-source LLMs, Mistral-7B, Llama-3-8B, vicuna-13b and OpenAI's API, gpt-4-turbo and gpt-4o-mini.}
% \label{open-source LLMs}
% \end{table}

Table \ref{open-source LLMs} in Appendix shows the performance of different listwise methods with different LLMs. The top-100 candidate documents are retrieved by BM25. The results of Setwise are still based on the code provided in the original paper of Setwise. Considering the cost and latency limitations of gpt-4-turbo and gpt-4o-mini APIs, we only run RankGPT as baseline, and TourRank-$r$ only run up to TourRank-5 (r=5).

In Table \ref{open-source LLMs}, TourRank-10 still performs very well on the three open-source models, and achieves state-of-the-art performance of in all of them except for TREC DL 20 based on vicuna. Even TourRank-2 sometimes show a relatively good performance. RankGPT's performance on the open-source models is not as good as the gpt-3.5-turbo API, but it still shows a stable and reasonable performance. However, Setwise performs poorly based on Llama3, especially on the TREC DL 20 dataset. In addition, TourRank-5 with gpt-4-turbo or gpt-4o-mini outperforms all methods based on gpt-3.5-turbo in Table \ref{TREC}, which indicates that TourRank can achieve higher performance with fewer tournaments times $r$ based on a stronger model. We also notice that RankGPT performs very well on the stronger models, gpt-4-turbo and gpt-4-mini, and RankGPT is even on par with TourRank-5 on TREC DL 20 based on these two APIs. This shows that RankGPT is also a very good zero-shot documents ranking method based on the strong base model.

In summary, TourRank and RankGPT are more stable and better performing zero-shot document ranking methods.

These experiments shows that TourRank can achieve good performance not only based on OpenAI's API, but also based on open-source LLMs, answering the \textbf{RQ.4}.

\section{Conclusions}

We introduce TourRank, a novel zero-shot document ranking method inspired by the tournament mechanism. TourRank allows for multiple tournaments to be conducted in parallel using a multi-process approach. This method effectively addresses challenges faced by large language models in ranking tasks, such as input length limitations, sensitivity to input order, and the difficulty of balancing effectiveness with efficiency.

Furthermore, our experiments indeed demonstrate that TourRank not only outperforms existing LLM-based zero-shot ranking methods but also successfully strikes a good balance between effectiveness and resource consumption. 

\section*{Acknowledgments}

This research was supported by the Natural Science Foundation of China (61902209, 62377044, U2001212), and Beijing Outstanding Young Scientist Program (NO.BJJWZYJH012019100020098), Intelligent Social Governance Platform, Major Innovation \& Planning Interdisciplinary Platform for the “Double-First Class” Initiative, Renmin University of China, Engineering Research Center of Next-Generation Intelligent Search and Recommendation, Ministry of Education, and the fund for building world-class universities (disciplines) of Renmin University of China.

% These results highlight the potential of TourRank not only in future zero-shot document ranking but also in the application of large language models for automated query-documents annotation.

% Bibliography entries for the entire Anthology, followed by custom entries
%\bibliography{anthology,custom}
% Custom bibliography entries only

% % \clearpage
% \section{Limitations}

% The performance of TourRank is inherently dependent on the capabilities of the underlying LLMs. If the LLMs can't follow the instructions well, it will be difficult to achieve good results. 

% Although multiple tournaments of TourRank can be performed in parallel in a multi-process manner, for example, based on the API of OpenAI, it is difficult to run the open-source models in a multi-process manner under the environment of limited computing resources.

% \bibliography{custom}

\clearpage
\bibliographystyle{ACM-Reference-Format}
\bibliography{sample-base}

%%% -*-BibTeX-*-
%%% Do NOT edit. File created by BibTeX with style
%%% ACM-Reference-Format-Journals [18-Jan-2012].

\begin{thebibliography}{38}

%%% ====================================================================
%%% NOTE TO THE USER: you can override these defaults by providing
%%% customized versions of any of these macros before the \bibliography
%%% command.  Each of them MUST provide its own final punctuation,
%%% except for \shownote{}, \showDOI{}, and \showURL{}.  The latter two
%%% do not use final punctuation, in order to avoid confusing it with
%%% the Web address.
%%%
%%% To suppress output of a particular field, define its macro to expand
%%% to an empty string, or better, \unskip, like this:
%%%
%%% \newcommand{\showDOI}[1]{\unskip}   % LaTeX syntax
%%%
%%% \def \showDOI #1{\unskip}           % plain TeX syntax
%%%
%%% ====================================================================

\ifx \showCODEN    \undefined \def \showCODEN     #1{\unskip}     \fi
\ifx \showDOI      \undefined \def \showDOI       #1{#1}\fi
\ifx \showISBNx    \undefined \def \showISBNx     #1{\unskip}     \fi
\ifx \showISBNxiii \undefined \def \showISBNxiii  #1{\unskip}     \fi
\ifx \showISSN     \undefined \def \showISSN      #1{\unskip}     \fi
\ifx \showLCCN     \undefined \def \showLCCN      #1{\unskip}     \fi
\ifx \shownote     \undefined \def \shownote      #1{#1}          \fi
\ifx \showarticletitle \undefined \def \showarticletitle #1{#1}   \fi
\ifx \showURL      \undefined \def \showURL       {\relax}        \fi
% The following commands are used for tagged output and should be
% invisible to TeX
\providecommand\bibfield[2]{#2}
\providecommand\bibinfo[2]{#2}
\providecommand\natexlab[1]{#1}
\providecommand\showeprint[2][]{arXiv:#2}

\bibitem[Achiam et~al\mbox{.}(2023)]%
        {achiam2023gpt}
\bibfield{author}{\bibinfo{person}{Josh Achiam}, \bibinfo{person}{Steven Adler}, \bibinfo{person}{Sandhini Agarwal}, \bibinfo{person}{Lama Ahmad}, \bibinfo{person}{Ilge Akkaya}, \bibinfo{person}{Florencia~Leoni Aleman}, \bibinfo{person}{Diogo Almeida}, \bibinfo{person}{Janko Altenschmidt}, \bibinfo{person}{Sam Altman}, \bibinfo{person}{Shyamal Anadkat}, {et~al\mbox{.}}} \bibinfo{year}{2023}\natexlab{}.
\newblock \showarticletitle{Gpt-4 technical report}.
\newblock \bibinfo{journal}{\emph{arXiv preprint arXiv:2303.08774}} (\bibinfo{year}{2023}).
\newblock


\bibitem[Bajaj et~al\mbox{.}(2016)]%
        {bajaj2016ms}
\bibfield{author}{\bibinfo{person}{Payal Bajaj}, \bibinfo{person}{Daniel Campos}, \bibinfo{person}{Nick Craswell}, \bibinfo{person}{Li Deng}, \bibinfo{person}{Jianfeng Gao}, \bibinfo{person}{Xiaodong Liu}, \bibinfo{person}{Rangan Majumder}, \bibinfo{person}{Andrew McNamara}, \bibinfo{person}{Bhaskar Mitra}, \bibinfo{person}{Tri Nguyen}, {et~al\mbox{.}}} \bibinfo{year}{2016}\natexlab{}.
\newblock \showarticletitle{Ms marco: A human generated machine reading comprehension dataset}.
\newblock \bibinfo{journal}{\emph{arXiv preprint arXiv:1611.09268}} (\bibinfo{year}{2016}).
\newblock


\bibitem[Craswell et~al\mbox{.}(2021)]%
        {craswell2021overview}
\bibfield{author}{\bibinfo{person}{Nick Craswell}, \bibinfo{person}{Bhaskar Mitra}, \bibinfo{person}{Emine Yilmaz}, {and} \bibinfo{person}{Daniel Campos}.} \bibinfo{year}{2021}\natexlab{}.
\newblock \bibinfo{title}{Overview of the TREC 2020 deep learning track}.
\newblock
\newblock
\showeprint[arxiv]{2102.07662}~[cs.IR]


\bibitem[Craswell et~al\mbox{.}(2020)]%
        {craswell2020overview}
\bibfield{author}{\bibinfo{person}{Nick Craswell}, \bibinfo{person}{Bhaskar Mitra}, \bibinfo{person}{Emine Yilmaz}, \bibinfo{person}{Daniel Campos}, {and} \bibinfo{person}{Ellen~M Voorhees}.} \bibinfo{year}{2020}\natexlab{}.
\newblock \showarticletitle{Overview of the TREC 2019 deep learning track}.
\newblock \bibinfo{journal}{\emph{arXiv preprint arXiv:2003.07820}} (\bibinfo{year}{2020}).
\newblock


\bibitem[Devlin et~al\mbox{.}(2018)]%
        {devlin2018bert}
\bibfield{author}{\bibinfo{person}{Jacob Devlin}, \bibinfo{person}{Ming-Wei Chang}, \bibinfo{person}{Kenton Lee}, {and} \bibinfo{person}{Kristina Toutanova}.} \bibinfo{year}{2018}\natexlab{}.
\newblock \showarticletitle{Bert: Pre-training of deep bidirectional transformers for language understanding}.
\newblock \bibinfo{journal}{\emph{arXiv preprint arXiv:1810.04805}} (\bibinfo{year}{2018}).
\newblock


\bibitem[Fan et~al\mbox{.}(2022)]%
        {fan2022pre}
\bibfield{author}{\bibinfo{person}{Yixing Fan}, \bibinfo{person}{Xiaohui Xie}, \bibinfo{person}{Yinqiong Cai}, \bibinfo{person}{Jia Chen}, \bibinfo{person}{Xinyu Ma}, \bibinfo{person}{Xiangsheng Li}, \bibinfo{person}{Ruqing Zhang}, \bibinfo{person}{Jiafeng Guo}, {et~al\mbox{.}}} \bibinfo{year}{2022}\natexlab{}.
\newblock \showarticletitle{Pre-training methods in information retrieval}.
\newblock \bibinfo{journal}{\emph{Foundations and Trends{\textregistered} in Information Retrieval}} \bibinfo{volume}{16}, \bibinfo{number}{3} (\bibinfo{year}{2022}), \bibinfo{pages}{178--317}.
\newblock


\bibitem[Formal et~al\mbox{.}(2022)]%
        {formal2022distillation}
\bibfield{author}{\bibinfo{person}{Thibault Formal}, \bibinfo{person}{Carlos Lassance}, \bibinfo{person}{Benjamin Piwowarski}, {and} \bibinfo{person}{St{\'e}phane Clinchant}.} \bibinfo{year}{2022}\natexlab{}.
\newblock \showarticletitle{From distillation to hard negative sampling: Making sparse neural ir models more effective}. In \bibinfo{booktitle}{\emph{Proceedings of the 45th international ACM SIGIR conference on research and development in information retrieval}}. \bibinfo{pages}{2353--2359}.
\newblock


\bibitem[Guo et~al\mbox{.}(2024)]%
        {guo2024generating}
\bibfield{author}{\bibinfo{person}{Fang Guo}, \bibinfo{person}{Wenyu Li}, \bibinfo{person}{Honglei Zhuang}, \bibinfo{person}{Yun Luo}, \bibinfo{person}{Yafu Li}, \bibinfo{person}{Le Yan}, {and} \bibinfo{person}{Yue Zhang}.} \bibinfo{year}{2024}\natexlab{}.
\newblock \showarticletitle{Generating Diverse Criteria On-the-Fly to Improve Point-wise LLM Rankers}.
\newblock \bibinfo{journal}{\emph{arXiv preprint arXiv:2404.11960}} (\bibinfo{year}{2024}).
\newblock


\bibitem[Izacard et~al\mbox{.}(2021)]%
        {izacard2021unsupervised}
\bibfield{author}{\bibinfo{person}{Gautier Izacard}, \bibinfo{person}{Mathilde Caron}, \bibinfo{person}{Lucas Hosseini}, \bibinfo{person}{Sebastian Riedel}, \bibinfo{person}{Piotr Bojanowski}, \bibinfo{person}{Armand Joulin}, {and} \bibinfo{person}{Edouard Grave}.} \bibinfo{year}{2021}\natexlab{}.
\newblock \showarticletitle{Unsupervised dense information retrieval with contrastive learning}.
\newblock \bibinfo{journal}{\emph{arXiv preprint arXiv:2112.09118}} (\bibinfo{year}{2021}).
\newblock


\bibitem[Jiang et~al\mbox{.}(2023)]%
        {jiang2023mistral}
\bibfield{author}{\bibinfo{person}{Albert~Q Jiang}, \bibinfo{person}{Alexandre Sablayrolles}, \bibinfo{person}{Arthur Mensch}, \bibinfo{person}{Chris Bamford}, \bibinfo{person}{Devendra~Singh Chaplot}, \bibinfo{person}{Diego de~las Casas}, \bibinfo{person}{Florian Bressand}, \bibinfo{person}{Gianna Lengyel}, \bibinfo{person}{Guillaume Lample}, \bibinfo{person}{Lucile Saulnier}, {et~al\mbox{.}}} \bibinfo{year}{2023}\natexlab{}.
\newblock \showarticletitle{Mistral 7B}.
\newblock \bibinfo{journal}{\emph{arXiv preprint arXiv:2310.06825}} (\bibinfo{year}{2023}).
\newblock


\bibitem[Liang et~al\mbox{.}(2022)]%
        {liang2022holistic}
\bibfield{author}{\bibinfo{person}{Percy Liang}, \bibinfo{person}{Rishi Bommasani}, \bibinfo{person}{Tony Lee}, \bibinfo{person}{Dimitris Tsipras}, \bibinfo{person}{Dilara Soylu}, \bibinfo{person}{Michihiro Yasunaga}, \bibinfo{person}{Yian Zhang}, \bibinfo{person}{Deepak Narayanan}, \bibinfo{person}{Yuhuai Wu}, \bibinfo{person}{Ananya Kumar}, {et~al\mbox{.}}} \bibinfo{year}{2022}\natexlab{}.
\newblock \showarticletitle{Holistic evaluation of language models}.
\newblock \bibinfo{journal}{\emph{arXiv preprint arXiv:2211.09110}} (\bibinfo{year}{2022}).
\newblock


\bibitem[Liu et~al\mbox{.}(2024a)]%
        {liu2024lost}
\bibfield{author}{\bibinfo{person}{Nelson~F Liu}, \bibinfo{person}{Kevin Lin}, \bibinfo{person}{John Hewitt}, \bibinfo{person}{Ashwin Paranjape}, \bibinfo{person}{Michele Bevilacqua}, \bibinfo{person}{Fabio Petroni}, {and} \bibinfo{person}{Percy Liang}.} \bibinfo{year}{2024}\natexlab{a}.
\newblock \showarticletitle{Lost in the middle: How language models use long contexts}.
\newblock \bibinfo{journal}{\emph{Transactions of the Association for Computational Linguistics}}  \bibinfo{volume}{12} (\bibinfo{year}{2024}), \bibinfo{pages}{157--173}.
\newblock


\bibitem[Liu et~al\mbox{.}(2024b)]%
        {liu2024sliding}
\bibfield{author}{\bibinfo{person}{Wenhan Liu}, \bibinfo{person}{Xinyu Ma}, \bibinfo{person}{Yutao Zhu}, \bibinfo{person}{Ziliang Zhao}, \bibinfo{person}{Shuaiqiang Wang}, \bibinfo{person}{Dawei Yin}, {and} \bibinfo{person}{Zhicheng Dou}.} \bibinfo{year}{2024}\natexlab{b}.
\newblock \showarticletitle{Sliding Windows Are Not the End: Exploring Full Ranking with Long-Context Large Language Models}.
\newblock \bibinfo{journal}{\emph{arXiv preprint arXiv:2412.14574}} (\bibinfo{year}{2024}).
\newblock


\bibitem[Liu et~al\mbox{.}(2024c)]%
        {liu2024demorank}
\bibfield{author}{\bibinfo{person}{Wenhan Liu}, \bibinfo{person}{Yutao Zhu}, {and} \bibinfo{person}{Zhicheng Dou}.} \bibinfo{year}{2024}\natexlab{c}.
\newblock \showarticletitle{Demorank: Selecting effective demonstrations for large language models in ranking task}.
\newblock \bibinfo{journal}{\emph{arXiv preprint arXiv:2406.16332}} (\bibinfo{year}{2024}).
\newblock


\bibitem[Luo et~al\mbox{.}(2024)]%
        {luo2024prp}
\bibfield{author}{\bibinfo{person}{Jian Luo}, \bibinfo{person}{Xuanang Chen}, \bibinfo{person}{Ben He}, {and} \bibinfo{person}{Le Sun}.} \bibinfo{year}{2024}\natexlab{}.
\newblock \showarticletitle{PRP-Graph: Pairwise Ranking Prompting to LLMs with Graph Aggregation for Effective Text Re-ranking}. In \bibinfo{booktitle}{\emph{Proceedings of the 62nd Annual Meeting of the Association for Computational Linguistics (Volume 1: Long Papers)}}. \bibinfo{pages}{5766--5776}.
\newblock


\bibitem[Ma et~al\mbox{.}(2023)]%
        {ma2023zero}
\bibfield{author}{\bibinfo{person}{Xueguang Ma}, \bibinfo{person}{Xinyu Zhang}, \bibinfo{person}{Ronak Pradeep}, {and} \bibinfo{person}{Jimmy Lin}.} \bibinfo{year}{2023}\natexlab{}.
\newblock \showarticletitle{Zero-shot listwise document reranking with a large language model}.
\newblock \bibinfo{journal}{\emph{arXiv preprint arXiv:2305.02156}} (\bibinfo{year}{2023}).
\newblock


\bibitem[MetaAI(2024)]%
        {Meta2024}
\bibfield{author}{\bibinfo{person}{MetaAI}.} \bibinfo{year}{2024}\natexlab{}.
\newblock \bibinfo{title}{Meta Llama 3}.
\newblock
\newblock
\urldef\tempurl%
\url{https://ai.meta.com/blog/meta-llama-3/}
\showURL{%
\tempurl}


\bibitem[Nogueira and Cho(2019)]%
        {nogueira2019passage}
\bibfield{author}{\bibinfo{person}{Rodrigo Nogueira} {and} \bibinfo{person}{Kyunghyun Cho}.} \bibinfo{year}{2019}\natexlab{}.
\newblock \showarticletitle{Passage Re-ranking with BERT}.
\newblock \bibinfo{journal}{\emph{arXiv preprint arXiv:1901.04085}} (\bibinfo{year}{2019}).
\newblock


\bibitem[Nogueira et~al\mbox{.}(2020)]%
        {nogueira2020document}
\bibfield{author}{\bibinfo{person}{Rodrigo Nogueira}, \bibinfo{person}{Zhiying Jiang}, {and} \bibinfo{person}{Jimmy Lin}.} \bibinfo{year}{2020}\natexlab{}.
\newblock \showarticletitle{Document ranking with a pretrained sequence-to-sequence model}.
\newblock \bibinfo{journal}{\emph{arXiv preprint arXiv:2003.06713}} (\bibinfo{year}{2020}).
\newblock


\bibitem[OpenAI({[n.\,d.]})]%
        {openai}
\bibfield{author}{\bibinfo{person}{OpenAI}.} \bibinfo{year}{[n.\,d.]}\natexlab{}.
\newblock \bibinfo{title}{OpenAI}.
\newblock \bibinfo{howpublished}{\url{https://www.openai.com/}}.
\newblock
\newblock
\shownote{Optional additional information}.


\bibitem[Pradeep et~al\mbox{.}(2021)]%
        {pradeep2021expando}
\bibfield{author}{\bibinfo{person}{Ronak Pradeep}, \bibinfo{person}{Rodrigo Nogueira}, {and} \bibinfo{person}{Jimmy Lin}.} \bibinfo{year}{2021}\natexlab{}.
\newblock \showarticletitle{The expando-mono-duo design pattern for text ranking with pretrained sequence-to-sequence models}.
\newblock \bibinfo{journal}{\emph{arXiv preprint arXiv:2101.05667}} (\bibinfo{year}{2021}).
\newblock


\bibitem[Pradeep et~al\mbox{.}(2023a)]%
        {pradeep2023rankvicuna}
\bibfield{author}{\bibinfo{person}{Ronak Pradeep}, \bibinfo{person}{Sahel Sharifymoghaddam}, {and} \bibinfo{person}{Jimmy Lin}.} \bibinfo{year}{2023}\natexlab{a}.
\newblock \showarticletitle{Rankvicuna: Zero-shot listwise document reranking with open-source large language models}.
\newblock \bibinfo{journal}{\emph{arXiv preprint arXiv:2309.15088}} (\bibinfo{year}{2023}).
\newblock


\bibitem[Pradeep et~al\mbox{.}(2023b)]%
        {pradeep2023rankzephyr}
\bibfield{author}{\bibinfo{person}{Ronak Pradeep}, \bibinfo{person}{Sahel Sharifymoghaddam}, {and} \bibinfo{person}{Jimmy Lin}.} \bibinfo{year}{2023}\natexlab{b}.
\newblock \showarticletitle{RankZephyr: Effective and Robust Zero-Shot Listwise Reranking is a Breeze!}
\newblock \bibinfo{journal}{\emph{arXiv preprint arXiv:2312.02724}} (\bibinfo{year}{2023}).
\newblock


\bibitem[Qin et~al\mbox{.}(2023)]%
        {qin2023large}
\bibfield{author}{\bibinfo{person}{Zhen Qin}, \bibinfo{person}{Rolf Jagerman}, \bibinfo{person}{Kai Hui}, \bibinfo{person}{Honglei Zhuang}, \bibinfo{person}{Junru Wu}, \bibinfo{person}{Jiaming Shen}, \bibinfo{person}{Tianqi Liu}, \bibinfo{person}{Jialu Liu}, \bibinfo{person}{Donald Metzler}, \bibinfo{person}{Xuanhui Wang}, {et~al\mbox{.}}} \bibinfo{year}{2023}\natexlab{}.
\newblock \showarticletitle{Large language models are effective text rankers with pairwise ranking prompting}.
\newblock \bibinfo{journal}{\emph{arXiv preprint arXiv:2306.17563}} (\bibinfo{year}{2023}).
\newblock


\bibitem[Raffel et~al\mbox{.}(2020)]%
        {raffel2020exploring}
\bibfield{author}{\bibinfo{person}{Colin Raffel}, \bibinfo{person}{Noam Shazeer}, \bibinfo{person}{Adam Roberts}, \bibinfo{person}{Katherine Lee}, \bibinfo{person}{Sharan Narang}, \bibinfo{person}{Michael Matena}, \bibinfo{person}{Yanqi Zhou}, \bibinfo{person}{Wei Li}, {and} \bibinfo{person}{Peter~J Liu}.} \bibinfo{year}{2020}\natexlab{}.
\newblock \showarticletitle{Exploring the limits of transfer learning with a unified text-to-text transformer}.
\newblock \bibinfo{journal}{\emph{Journal of machine learning research}} \bibinfo{volume}{21}, \bibinfo{number}{140} (\bibinfo{year}{2020}), \bibinfo{pages}{1--67}.
\newblock


\bibitem[Robertson et~al\mbox{.}(2009)]%
        {robertson2009probabilistic}
\bibfield{author}{\bibinfo{person}{Stephen Robertson}, \bibinfo{person}{Hugo Zaragoza}, {et~al\mbox{.}}} \bibinfo{year}{2009}\natexlab{}.
\newblock \showarticletitle{The probabilistic relevance framework: BM25 and beyond}.
\newblock \bibinfo{journal}{\emph{Foundations and Trends{\textregistered} in Information Retrieval}} \bibinfo{volume}{3}, \bibinfo{number}{4} (\bibinfo{year}{2009}), \bibinfo{pages}{333--389}.
\newblock


\bibitem[Sachan et~al\mbox{.}(2022)]%
        {sachan2022improving}
\bibfield{author}{\bibinfo{person}{Devendra~Singh Sachan}, \bibinfo{person}{Mike Lewis}, \bibinfo{person}{Mandar Joshi}, \bibinfo{person}{Armen Aghajanyan}, \bibinfo{person}{Wen-tau Yih}, \bibinfo{person}{Joelle Pineau}, {and} \bibinfo{person}{Luke Zettlemoyer}.} \bibinfo{year}{2022}\natexlab{}.
\newblock \showarticletitle{Improving passage retrieval with zero-shot question generation}.
\newblock \bibinfo{journal}{\emph{arXiv preprint arXiv:2204.07496}} (\bibinfo{year}{2022}).
\newblock


\bibitem[Sun et~al\mbox{.}(2023a)]%
        {sun2023instruction}
\bibfield{author}{\bibinfo{person}{Weiwei Sun}, \bibinfo{person}{Zheng Chen}, \bibinfo{person}{Xinyu Ma}, \bibinfo{person}{Lingyong Yan}, \bibinfo{person}{Shuaiqiang Wang}, \bibinfo{person}{Pengjie Ren}, \bibinfo{person}{Zhumin Chen}, \bibinfo{person}{Dawei Yin}, {and} \bibinfo{person}{Zhaochun Ren}.} \bibinfo{year}{2023}\natexlab{a}.
\newblock \showarticletitle{Instruction distillation makes large language models efficient zero-shot rankers}.
\newblock \bibinfo{journal}{\emph{arXiv preprint arXiv:2311.01555}} (\bibinfo{year}{2023}).
\newblock


\bibitem[Sun et~al\mbox{.}(2023b)]%
        {sun2023chatgpt}
\bibfield{author}{\bibinfo{person}{Weiwei Sun}, \bibinfo{person}{Lingyong Yan}, \bibinfo{person}{Xinyu Ma}, \bibinfo{person}{Pengjie Ren}, \bibinfo{person}{Dawei Yin}, {and} \bibinfo{person}{Zhaochun Ren}.} \bibinfo{year}{2023}\natexlab{b}.
\newblock \showarticletitle{Is chatgpt good at search? investigating large language models as re-ranking agent}.
\newblock \bibinfo{journal}{\emph{arXiv preprint arXiv:2304.09542}} (\bibinfo{year}{2023}).
\newblock


\bibitem[Tang et~al\mbox{.}(2023)]%
        {tang2023found}
\bibfield{author}{\bibinfo{person}{Raphael Tang}, \bibinfo{person}{Xinyu Zhang}, \bibinfo{person}{Xueguang Ma}, \bibinfo{person}{Jimmy Lin}, {and} \bibinfo{person}{Ferhan Ture}.} \bibinfo{year}{2023}\natexlab{}.
\newblock \showarticletitle{Found in the middle: Permutation self-consistency improves listwise ranking in large language models}.
\newblock \bibinfo{journal}{\emph{arXiv preprint arXiv:2310.07712}} (\bibinfo{year}{2023}).
\newblock


\bibitem[Thakur et~al\mbox{.}(2024)]%
        {thakur2024systematic}
\bibfield{author}{\bibinfo{person}{Nandan Thakur}, \bibinfo{person}{Luiz Bonifacio}, \bibinfo{person}{Maik Fr{\"o}be}, \bibinfo{person}{Alexander Bondarenko}, \bibinfo{person}{Ehsan Kamalloo}, \bibinfo{person}{Martin Potthast}, \bibinfo{person}{Matthias Hagen}, {and} \bibinfo{person}{Jimmy Lin}.} \bibinfo{year}{2024}\natexlab{}.
\newblock \showarticletitle{Systematic Evaluation of Neural Retrieval Models on the Touch{\'e} 2020 Argument Retrieval Subset of BEIR}. In \bibinfo{booktitle}{\emph{Proceedings of the 47th International ACM SIGIR Conference on Research and Development in Information Retrieval}}.
\newblock


\bibitem[Thakur et~al\mbox{.}(2021)]%
        {thakur2021beir}
\bibfield{author}{\bibinfo{person}{Nandan Thakur}, \bibinfo{person}{Nils Reimers}, \bibinfo{person}{Andreas R{\"u}ckl{\'e}}, \bibinfo{person}{Abhishek Srivastava}, {and} \bibinfo{person}{Iryna Gurevych}.} \bibinfo{year}{2021}\natexlab{}.
\newblock \showarticletitle{Beir: A heterogenous benchmark for zero-shot evaluation of information retrieval models}.
\newblock \bibinfo{journal}{\emph{arXiv preprint arXiv:2104.08663}} (\bibinfo{year}{2021}).
\newblock


\bibitem[Yoon et~al\mbox{.}(2024)]%
        {yoon2024listt5}
\bibfield{author}{\bibinfo{person}{Soyoung Yoon}, \bibinfo{person}{Eunbi Lee}, \bibinfo{person}{Jiyeon Kim}, \bibinfo{person}{Yireun Kim}, \bibinfo{person}{Hyeongu Yun}, {and} \bibinfo{person}{Seung-won Hwang}.} \bibinfo{year}{2024}\natexlab{}.
\newblock \showarticletitle{ListT5: Listwise Reranking with Fusion-in-Decoder Improves Zero-shot Retrieval}.
\newblock \bibinfo{journal}{\emph{arXiv preprint arXiv:2402.15838}} (\bibinfo{year}{2024}).
\newblock


\bibitem[Zheng et~al\mbox{.}(2023)]%
        {zheng2023judging}
\bibfield{author}{\bibinfo{person}{Lianmin Zheng}, \bibinfo{person}{Wei-Lin Chiang}, \bibinfo{person}{Ying Sheng}, \bibinfo{person}{Siyuan Zhuang}, \bibinfo{person}{Zhanghao Wu}, \bibinfo{person}{Yonghao Zhuang}, \bibinfo{person}{Zi Lin}, \bibinfo{person}{Zhuohan Li}, \bibinfo{person}{Dacheng Li}, \bibinfo{person}{Eric Xing}, {et~al\mbox{.}}} \bibinfo{year}{2023}\natexlab{}.
\newblock \showarticletitle{Judging llm-as-a-judge with mt-bench and chatbot arena}.
\newblock \bibinfo{journal}{\emph{Advances in Neural Information Processing Systems}}  \bibinfo{volume}{36} (\bibinfo{year}{2023}), \bibinfo{pages}{46595--46623}.
\newblock


\bibitem[Zhu et~al\mbox{.}(2023)]%
        {zhu2023large}
\bibfield{author}{\bibinfo{person}{Yutao Zhu}, \bibinfo{person}{Huaying Yuan}, \bibinfo{person}{Shuting Wang}, \bibinfo{person}{Jiongnan Liu}, \bibinfo{person}{Wenhan Liu}, \bibinfo{person}{Chenlong Deng}, \bibinfo{person}{Haonan Chen}, \bibinfo{person}{Zheng Liu}, \bibinfo{person}{Zhicheng Dou}, {and} \bibinfo{person}{Ji-Rong Wen}.} \bibinfo{year}{2023}\natexlab{}.
\newblock \showarticletitle{Large language models for information retrieval: A survey}.
\newblock \bibinfo{journal}{\emph{arXiv preprint arXiv:2308.07107}} (\bibinfo{year}{2023}).
\newblock


\bibitem[Zhuang et~al\mbox{.}(2023a)]%
        {zhuang2023beyond}
\bibfield{author}{\bibinfo{person}{Honglei Zhuang}, \bibinfo{person}{Zhen Qin}, \bibinfo{person}{Kai Hui}, \bibinfo{person}{Junru Wu}, \bibinfo{person}{Le Yan}, \bibinfo{person}{Xuanhui Wang}, {and} \bibinfo{person}{Michael Berdersky}.} \bibinfo{year}{2023}\natexlab{a}.
\newblock \showarticletitle{Beyond yes and no: Improving zero-shot llm rankers via scoring fine-grained relevance labels}.
\newblock \bibinfo{journal}{\emph{arXiv preprint arXiv:2310.14122}} (\bibinfo{year}{2023}).
\newblock


\bibitem[Zhuang et~al\mbox{.}(2023b)]%
        {zhuang2023rankt5}
\bibfield{author}{\bibinfo{person}{Honglei Zhuang}, \bibinfo{person}{Zhen Qin}, \bibinfo{person}{Rolf Jagerman}, \bibinfo{person}{Kai Hui}, \bibinfo{person}{Ji Ma}, \bibinfo{person}{Jing Lu}, \bibinfo{person}{Jianmo Ni}, \bibinfo{person}{Xuanhui Wang}, {and} \bibinfo{person}{Michael Bendersky}.} \bibinfo{year}{2023}\natexlab{b}.
\newblock \showarticletitle{Rankt5: Fine-tuning t5 for text ranking with ranking losses}. In \bibinfo{booktitle}{\emph{Proceedings of the 46th International ACM SIGIR Conference on Research and Development in Information Retrieval}}. \bibinfo{pages}{2308--2313}.
\newblock


\bibitem[Zhuang et~al\mbox{.}(2023c)]%
        {zhuang2023setwise}
\bibfield{author}{\bibinfo{person}{Shengyao Zhuang}, \bibinfo{person}{Honglei Zhuang}, \bibinfo{person}{Bevan Koopman}, {and} \bibinfo{person}{Guido Zuccon}.} \bibinfo{year}{2023}\natexlab{c}.
\newblock \showarticletitle{A setwise approach for effective and highly efficient zero-shot ranking with large language models}.
\newblock \bibinfo{journal}{\emph{arXiv preprint arXiv:2310.09497}} (\bibinfo{year}{2023}).
\newblock


\end{thebibliography}

\appendix
% \clearpage
% \onecolumn 
\section*{Appendix}

\section{Related Works}
\label{More Related Works}

\subsection{Neural Network Approaches}

Documents ranking has made significant progress, with the help of pre-trained language models, such as BERT \cite{devlin2018bert} and T5 \cite{raffel2020exploring}. \citet{nogueira2019passage} present a multi-stage text ranking system using BERT, introducing monoBERT and duoBERT models that offer a balance between quality and latency, achieving state-of-the-art results on MS MARCO and TREC CAR datasets. \citet{nogueira2020document} introduce a new method for document ranking using a pre-trained sequence-to-sequence model, T5, which outperforms classification-based models, especially in data-poor scenarios, and demonstrates the model's ability to leverage latent knowledge from pretraining for improved performance. \citet{zhuang2023rankt5} introduce RankT5, a method for fine-tuning the T5 model for text ranking using ranking losses, which shows significant performance improvements over models fine-tuned with classification losses and demonstrates better zero-shot ranking performance on out-of-domain data.

\subsection{LLMs Approaches}
\label{LLMs Approaches}
\noindent
\textbf{Pointwise Approaches} \quad There are several works that employ various zero-shot pointwise rankers. Query Generation (QG) \cite{sachan2022improving} involves rescoring retrieved passages by leveraging a zero-shot question generation model. The model uses a pre-trained language model to compute the probability of the input question, conditioned on a retrieved passage. Binary Relevance Generation (B-RG) \cite{liang2022holistic} proposes to utilize LLMs to make predictions on a query-passage pair, utilizing the likelihood of ``Yes/No'' responses for the computation of ranking scores. The Rating Scale $0-k$ Relevance Generation (RS-RG) \cite{zhuang2023beyond} incorporates fine-grained relevance labels into the prompts for LLM rankers to better differentiate documents of varying relevance levels to the query, thereby achieving more accurate rankings. \citet{guo2024generating} propose a multi-perspective evaluation criteria-based ranking model to overcome the deficiencies of LLM rankers in standardized comparison and handling complex passages, thereby significantly enhancing the pointwise ranking performance. \citet{guo2024generating} have also considered the Rating Scale $0-k$ Directly Score (DIRECT(0, k)) method. This approach prompts the LLM to directly generate the relevance score for each query-passage pair. 

\noindent
\textbf{Pairwise Approaches} \quad \citet{pradeep2021expando} design a pairwise component to enhance the early precision performance of the text ranking system by employing a pre-trained sequence-to-sequence model (such as T5 \cite{raffel2020exploring}) to conduct pairwise comparisons and reranking of retrieved document pairs. \citet{qin2023large} introduce a method called Pairwise Ranking Prompting (PRP), which effectively enables LLMs to perform text ranking tasks by simplifying the prompt design and achieving competitive performance across multiple benchmark datasets. 

\noindent
\textbf{Listwise Approaches} \quad LRL \cite{ma2023zero} enhances text retrieval reranking by employing a large language model as a zero-shot listwise reranker, utilizing a simple instruction template and a sliding window strategy to process multi-document information. Similarly, \citet{sun2023chatgpt} introduce a novel instructional permutation generation approach called RankGPT, utilizing a sliding window strategy to effectively enable LLMs (such as ChatGPT \cite{openai} and GPT-4 \cite{achiam2023gpt}) to be used for relevance ranking tasks in information retrieval, achieving competitive and even superior results on popular IR benchmarks. In addition, both RankVicuna \cite{pradeep2023rankvicuna} and RankZephyr \cite{pradeep2023rankzephyr} utilize open-source LLMs and employ instruction fine-tuning to achieve zero-shot listwise document reranking, thereby enhancing the ranking performance of smaller LLMs. \citet{zhuang2023setwise} propose a novel Setwise prompting approach to enhance the efficiency and effectiveness of LLMs in zero-shot document ranking tasks, by reducing the number of model inferences and prompt token consumption, which significantly improves computational efficiency while maintaining high ranking performance. \citet{tang2023found} introduces permutation self-consistency, a method to reduce positional bias in large language models for listwise ranking tasks, achieving state-of-the-art performance in passage reranking and sorting datasets. DemoRank \cite{liu2024demorank} is a listwise ranking method to select and optimize the order of in-context demonstrations for large language models, improving their performance in few-shot learning tasks. \citet{liu2024sliding} explores full ranking with long-context LLMs, proposing a new listwise training approach and importance-aware learning objective to improve efficiency and performance in ranking tasks. \citet{yoon2024listt5} introduces ListT5, a listwise reranking model that utilizes Fusion-in-Decoder architecture for efficient and robust zero-shot retrieval, outperforming state-of-the-art methods on the BEIR benchmark. 

Since both ListT5 \cite{yoon2024listt5} and our TourRank both refer to the concept of tournament, it's worth clarifying the difference between two methods: 
(1) ListT5 is trained on the MS MARCO dataset \cite{bajaj2016ms}, while TourRank is a completely zero-shot method.
(2) The model of ListT5 is trained from an open-source model with encoder-decoder architecture to achieve good ranking efficiency, while TourRank can be applied to both open-source and closed-source models.
(3) Although both of two methods propose the basic unit of ranking or sorting inspired by tournament mechanism, the tournament in these two methods are different. In ListT5, the tournament selects the most relevant document. In TourRank, the tournament selects a few most relevant documents, and different stages correspond to different accumulated points $P_{T_r}$ (Table \ref{points}).
(4) In ListT5, to select the top-k most relevant documents, k times tournaments should be performed sequentially. TourRank, on the other hand, has an independent accumulated points system (Table \ref{points}) for each tournament, and multiple tournaments can be executed in parallel, which can significantly speed up the ranking process. The independent accumulated points system is also the key for TourRank to performing multiple tournaments in parallel.

\begin{figure*}[t]
\centering
\begin{subfigure}{0.495\textwidth}
  \centering
  \includegraphics[width=1\linewidth]{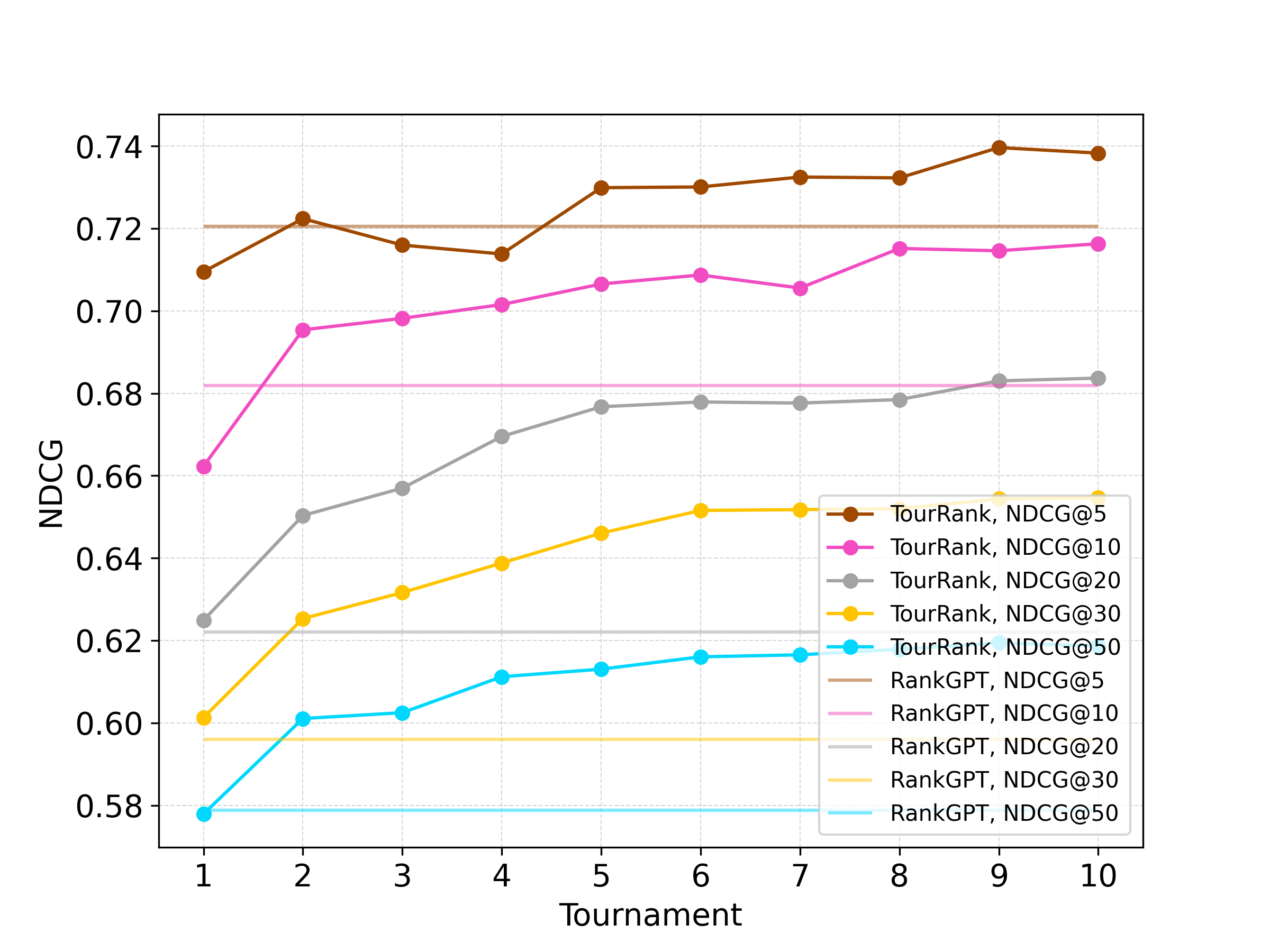}
  \caption{TREC DL 19}
  \label{trade_off_dl19}
\end{subfigure}
% \vspace{1cm} % 这里1cm是你想要的距离，你可以更改这个数值
\begin{subfigure}{0.495\textwidth}
  \centering
  \includegraphics[width=1\linewidth]{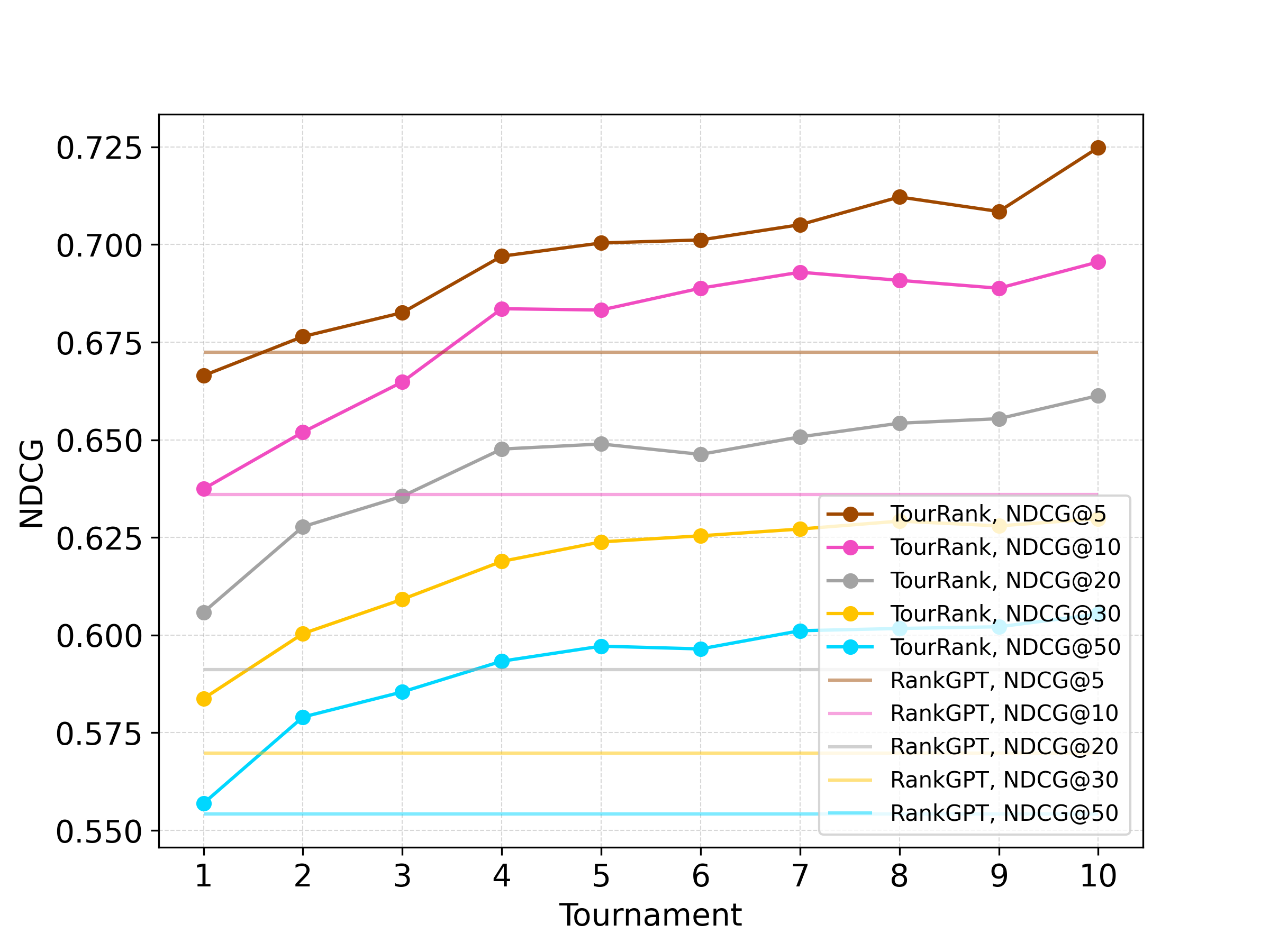}
  \caption{TREC DL 20}
  \label{trade_off_dl20}
\end{subfigure}
\begin{subfigure}{0.495\textwidth}
  \centering
  \includegraphics[width=1\linewidth]{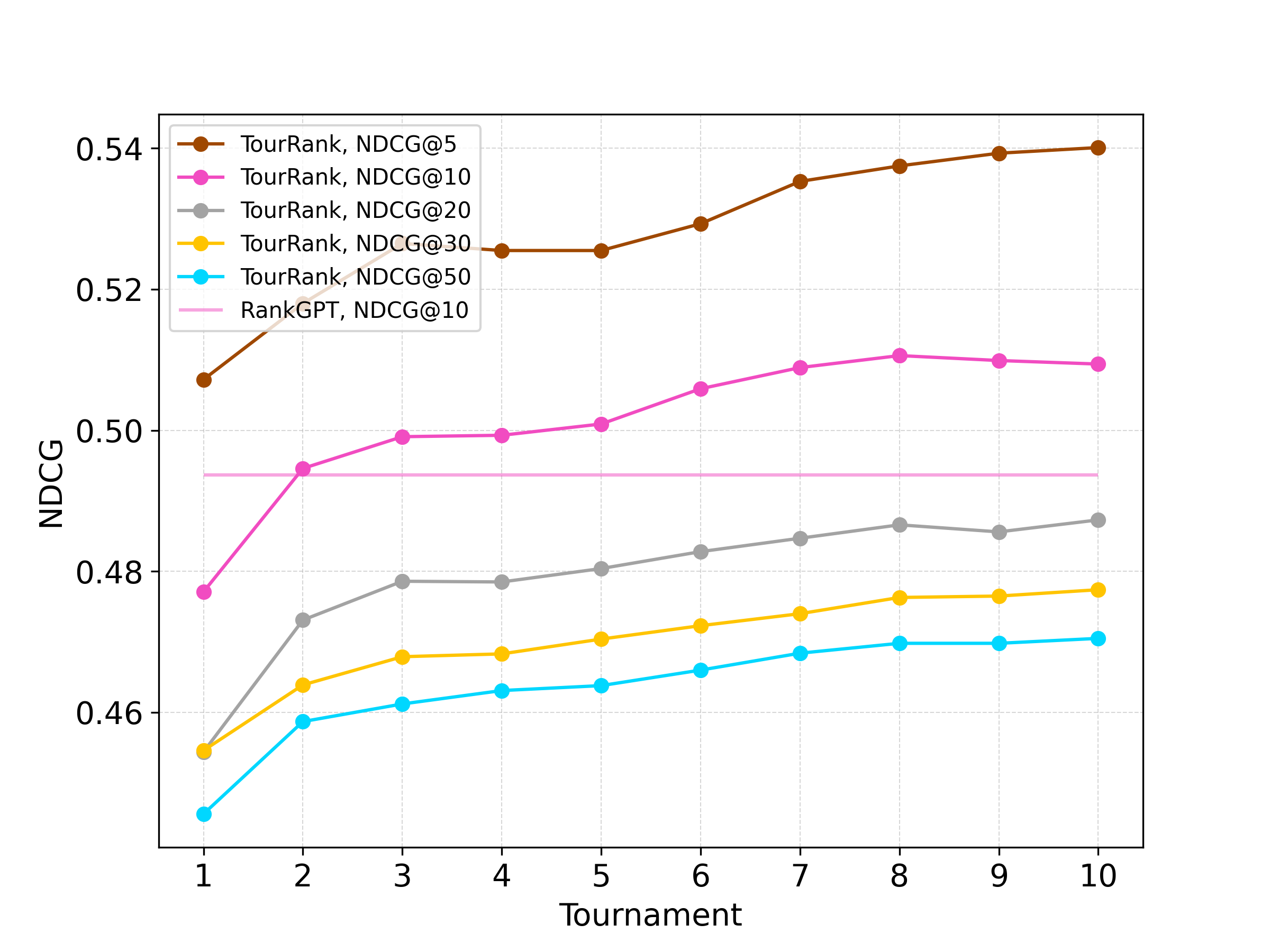}
  \caption{Average of 8 tasks on BEIR}
  \label{trade_off_beir}
\end{subfigure}
\caption{The performance of TourRank with different times of tournaments. The abscissa is the times of tournaments, and the ordinate is NDCG@\{5, 10, 20, 30, 50\}. All the results are based on gpt-3.5-turbo API.}
\label{trade_off}
\end{figure*}

\begin{figure*}[th]
\centering
\begin{subfigure}{0.495\textwidth}
  \centering
  \includegraphics[width=1\linewidth]{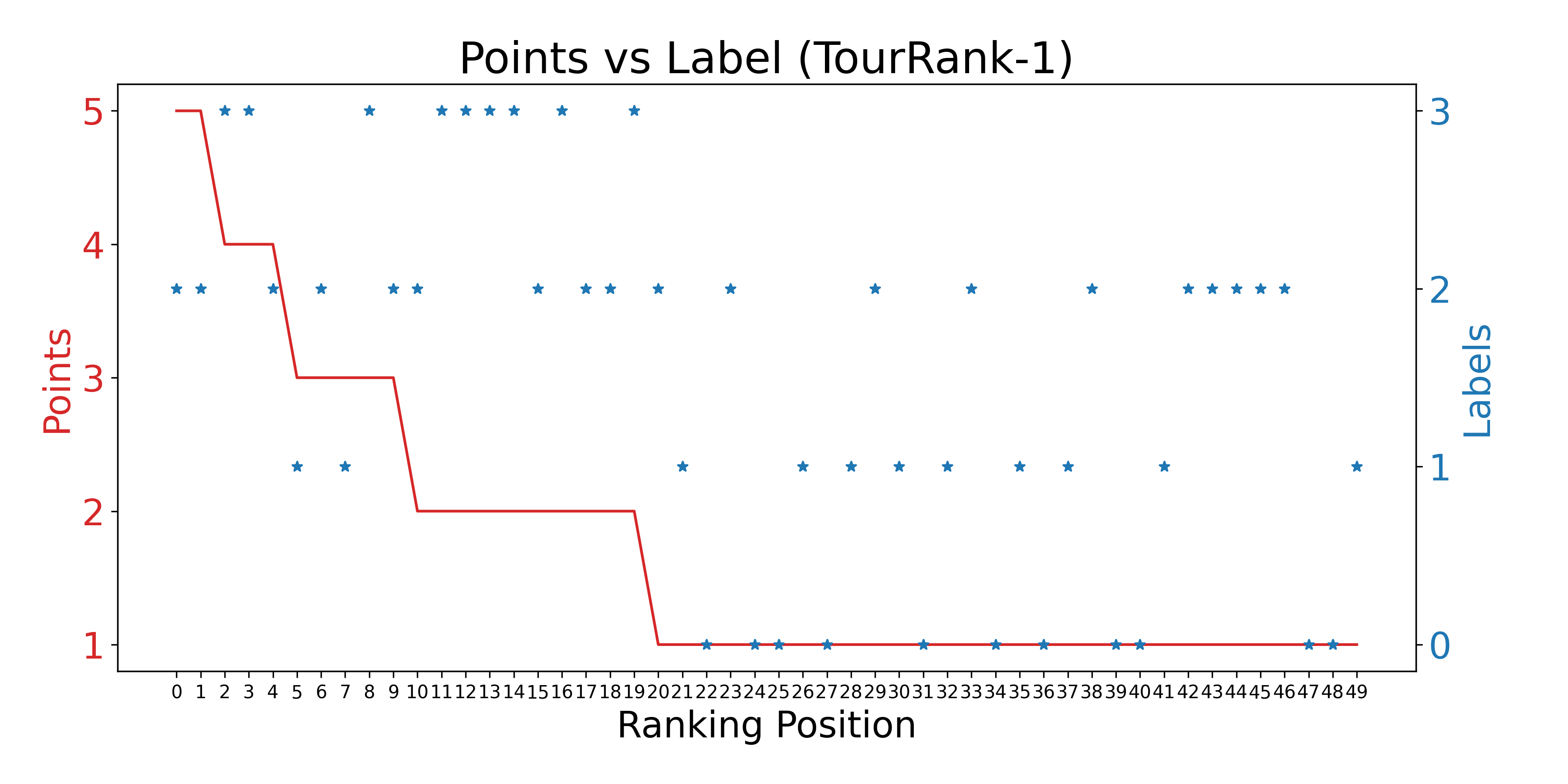}
  % \caption{TREC DL 19}
  % \label{sensitivity_dl19}
\end{subfigure}
% \vspace{1cm} % 这里1cm是你想要的距离，你可以更改这个数值
% \begin{subfigure}{0.5\textwidth}
%   \centering
%   \includegraphics[width=0.95\linewidth]{fig/index_vs_label_0_5.png}
%   % \caption{TREC DL 20}
%   % \label{sensitivity_dl20}
% \end{subfigure}
\begin{subfigure}{0.495\textwidth}
  \centering
  \includegraphics[width=1\linewidth]{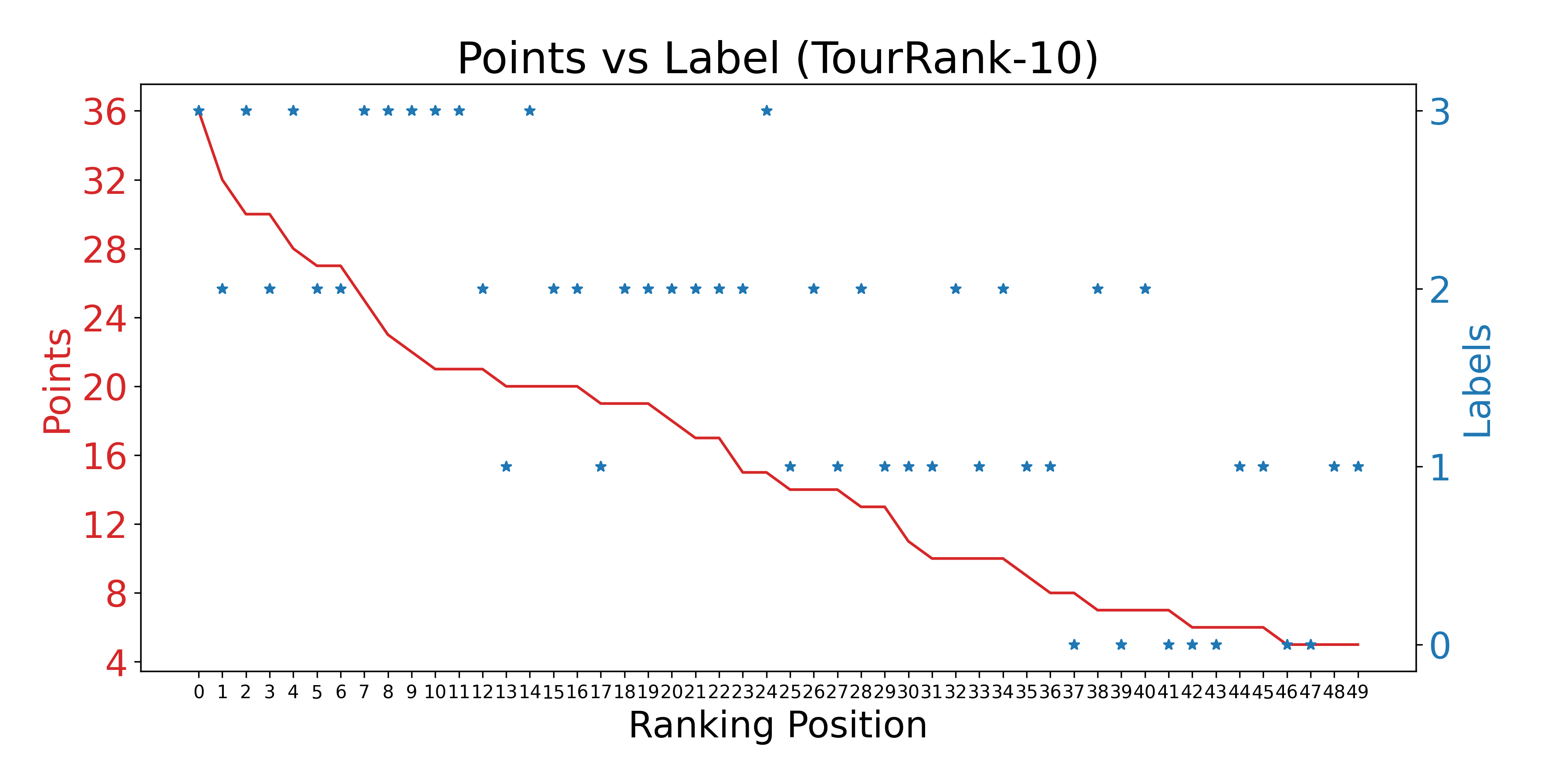}
  % \caption{TREC DL 20}
  % \label{sensitivity_dl20}
\end{subfigure}
\caption{The relationship between the accumulated points $P_T$ and the corresponding labels for TourRank-1 and TourRank-10. The query of this case is "how long is life cycle of flea" which is one of the queries in the TREC DL 19.}
\label{points_vs_labels}
\end{figure*}

\section{The Performance of TourRank-$r$}

Figure \ref{trade_off} shows the trend of NDCG@\{5, 10, 20, 30, 50\} with the increase of the number of tournaments for TourRank on TREC datasets and BEIR benchmark. We can see that after the first two tournaments, TourRank-2 achieves relatively good results on all datasets, outperforming RankGPT on all corresponding metrics shown. Even in TourRank-10, the metrics still have the potential to continue to increase. 

Since the number of tokens consumed by TourRank scales linearly with the number of tournaments, we can control the number of consumed tokens by controlling the number of tournaments. Thus, the balance between effectiveness and token consumption can be achieved. 

\section{Case Study: How Does TourRank Improve the Performance of Documents Ranking?}
\label{Case Study: How Does TourRank Improve the Performance of Documents Ranking?}

In Figure \ref{points_vs_labels}, the horizontal coordinate represents the ranking position of top-50 documents, the red lines represent the accumulated points $P_T$ of TourRank-1 and TourRank-10 respectively, and the blue star points represent the corresponding real labels (integers from 0 to 3). 

It can be seen that the $P_T$ of TourRank-1 is coarse, and the labels for the top-50 ranked documents are also relatively scattered. However, the accumulated points $P_T$ of TourRank-10 become much more fine-grained after 10 tournaments, and the labels corresponding to top-50 documents are relatively concentrated. After testing, the NDCG@\{10, 50\} of the case query have increased from \{0.7078, 0.8186\} to \{0.8715, 0.911\}. 

Therefore, as the times of tournaments increases, the accumulated points $P_T$ become more fine-grained. This is how exactly TourRank improves the document ranking performance.

\section{The Discussions on Time Complexity and Number of Documents Inputted to LLMs}
\label{The Discussions on Time Complexity and Number of Documents Inputted to LLMs}

Table \ref{time_token_2} is a more precise version of Table \ref{time_token} which shows the theoretical lowest time complexity of various methods and the number of documents which are inputted to LLMs for each method. Then, we analysis the content in Table \ref{time_token_2}.

\subsection{Time Complexity}

\noindent
\textbf{PointWise and Pairwise} \quad Since PointWise scoring a single document and PRP-Allpair comparing a pair of documents can be performed in parallel, the theoretical lowest time complexity is $O(1)$. However, since pairwise methods need to compare about $O(N^2)$ pairs of documents, the theoretical minimum time complexity $O(1)$ is difficult to implement.

\noindent
\textbf{Setwise.bubblesort} \quad According to \cite{zhuang2023setwise}, the time complexity of Setwise.bubblesort is $O(k * \frac{N}{c-1})$. Setwise rank the top-$k$ ($k<N$) documents through bubblesort, and $c$ is the documents compared in a prompt of Setwise. Considering that Setwise can achieve the best performance with $c=10$ based on gpt-3.5-turbo API, the time complexity is:

\begin{align*}
O(k * \frac{N}{c-1}) \approx O(\frac{1}{9}k*N)
\end{align*}

\noindent
\textbf{RankGPT} \quad RankGPT uses sliding window strategy, so its time complexity is $O(\frac{N-\omega}{s})$. The best window size is $\omega=20$ and the best step size is $s=10$ in RankGPT. Based on the optimal parameters ($\omega=20$ and $s=10$) and considering that $\omega$ is often much smaller than $N$, the best time complexity of RankGPT is:

\begin{align*}
O(\frac{N-\omega}{s}) = O(\frac{N-20}{10}) \approx O(\frac{1}{10}*N)
\end{align*}

\noindent
\textbf{TourRank-$r$} \quad One tournament includes $K-1$ times selection stages shown in Figure \ref{tournament_grouping}, so the time complexity of one tournament is $O(K-1)$. Because $r$ rounds tournaments can be performed in parallel, the time complexity of TourRank-$r$ is also $O(K-1)$.

\begin{table*}[h] %\tiny %\normalsize
\centering
\resizebox{0.62\textwidth}{!}{%
\begin{tabular}{ccc}
\toprule
% \hline
\textbf{Methods} & \textbf{Time Complexity} & \textbf{No. of Docs to LLMs} \\ 
\midrule
PointWise & $O(1)$ & $N$ \\ %\hline
\midrule
PRP-Allpair & $O(1)$ & $N^2-N$ \\ %\hline
\midrule
Setwise.bubblesort & $O(k * \frac{N}{c-1}) \approx O(\frac{1}{9} k*N)$ & $ k * \frac{N}{c-1} * c \approx \frac{10}{9} k * N$ \\ %\hline
\midrule
Setwise.heapsort & $O(k * log_c N) \approx O(k * log_{10} N)$ & $ k * log_c N * c \approx 10k * log_{10} N$ \\ %\hline
\midrule
RankGPT & $O(\frac{N-\omega}{s}) \approx O(\frac{1}{10}*N)$ & $\omega * \frac{N-\omega}{s} \approx 2*N$ \\ %\hline
\midrule
TourRank-$r$ & $O(K-1)$ & $\left( \sum_{k=0}^{K-1} \frac{N}{2^k} \right)*r \approx 2r*N$ \\ 
\bottomrule
\end{tabular}%
}
\caption{This Table is a more precise version of Table \ref{time_token}. The theoretical lowest time complexity of various methods and the number of documents which are inputted to LLMs for each method. $N$ is the number of candidate documents. Setwise ranks the top-$k$ ($k<N$) documents through bubblesort and heapsort, and $c=10$ is the documents compared in a prompt of Setwise based on gpt-3.5-turbo API. $\omega=20$ is window size and $s=10$ is step size in RankGPT. $K-1$ is the times of the selection stages in a tournament (Figure \ref{tournament_grouping} (a)) and $r$ is the times of tournaments in TourRank-$r$. All the approximate contents in this table are based on the recommended parameters.}
% \caption{The theoretical lowest time complexity of various methods and the number of documents which are inputted to LLMs for each method. $N$ is the number of candidate documents. $K-1$ is the times of the selection stage in a tournament (Figure \ref{tournament_grouping}). $\omega$ is window size and $s$ is step size in RankGPT. $r$ is the times of tournaments in TourRank.}
\label{time_token_2}
\end{table*}

\subsection{No. of Docs to LLMs}

\noindent
\textbf{PointWise} \quad Since the PointWise method scores each document once, the number of documents inputted to LLMs is $N$. 

\noindent
\textbf{Pairwise} \quad However, PRP-Allpair needs to form at least $\frac{N*(N-1)}{2}$ pairs for $N$ candidate documents, and since one pair of documents is inputted to LLMs each time, the number of documents it inputs to LLMs is $N^2-N$. 

\noindent
\textbf{Setwise.bubblesort} \quad The time complexity of Setwise.bubblesort is $O(k * \frac{N}{c-1})$ and $c=10$ documents is compared in a prompt, so the number of documents inputted to LLMs for Setwise.bubblesort is:

\begin{align*}
k * \frac{N}{c-1} * c \approx \frac{10}{9} k * N
\end{align*}

\noindent
\textbf{RankGPT} \quad In RankGPT, we know that $\omega$ documents need to be inputted into each window, and a total $\frac{N-\omega}{s}$ intra-window ranking need to be performed, so the number of documents input to LLMs is $\omega * \frac{N-\omega}{s}$. The best window size $\omega$ given in RankGPT is 20 and the best step size $s$ is 10. Based on the optimal parameters and considering that $\omega$ is often much smaller than $N$, the number of documents inputted into the LLMs of RankGPT is:

\begin{align*}
\omega * \frac{N-\omega}{s} = 20 * \frac{N-20}{10} \approx 2*N
\end{align*}

\noindent
\textbf{TourRank-$r$} \quad In TourRank, if close to half of the documents are selected to advance to the next selection stage in a tournament (that is, $m \approx \frac{1}{2}n$), the total number of documents input to LLMs is about:

\begin{align*}
N + \frac{N}{2} + \cdots + \frac{N}{2^{K-2}} &= \sum_{k=0}^{K-1} \frac{N}{2^k} \\
&= N * \frac{1-\left(\frac{1}{2}\right)^{K-1}}{1-\frac{1}{2}} \\
&\approx 2*N
\end{align*}

The TourRank-$r$ performs $r$ rounds tournaments, so the number of documents inputted to LLMs of TourRank is about:

\begin{align*}
\left( \sum_{k=0}^{K-1} \frac{N}{2^k} \right) * r \approx 2r * N
\end{align*}

\section{Comparison Between Serial RankGPT and Parallel TourRank-$r$}
\label{Comparison Between Serial RankGPT and Parallel TourRank-$r$}

\begin{figure*}[t]
\centering
\begin{subfigure}{0.495\textwidth}
  \centering
  \includegraphics[width=1\linewidth]{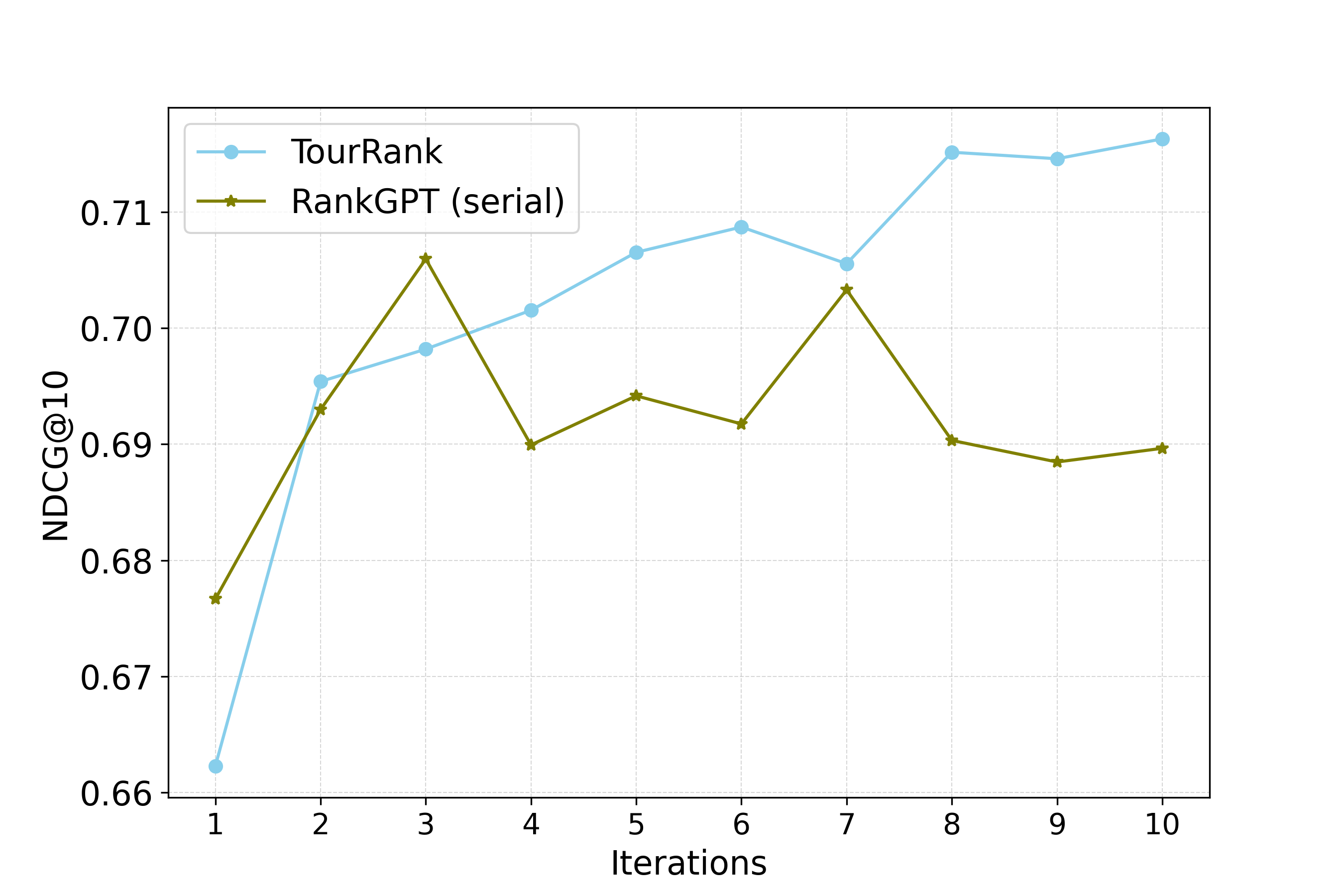}
  \caption{TREC DL 19}
  \label{chuanxing_dl19}
\end{subfigure}
% \vspace{1cm} % 这里1cm是你想要的距离，你可以更改这个数值
\begin{subfigure}{0.495\textwidth}
  \centering
  \includegraphics[width=1\linewidth]{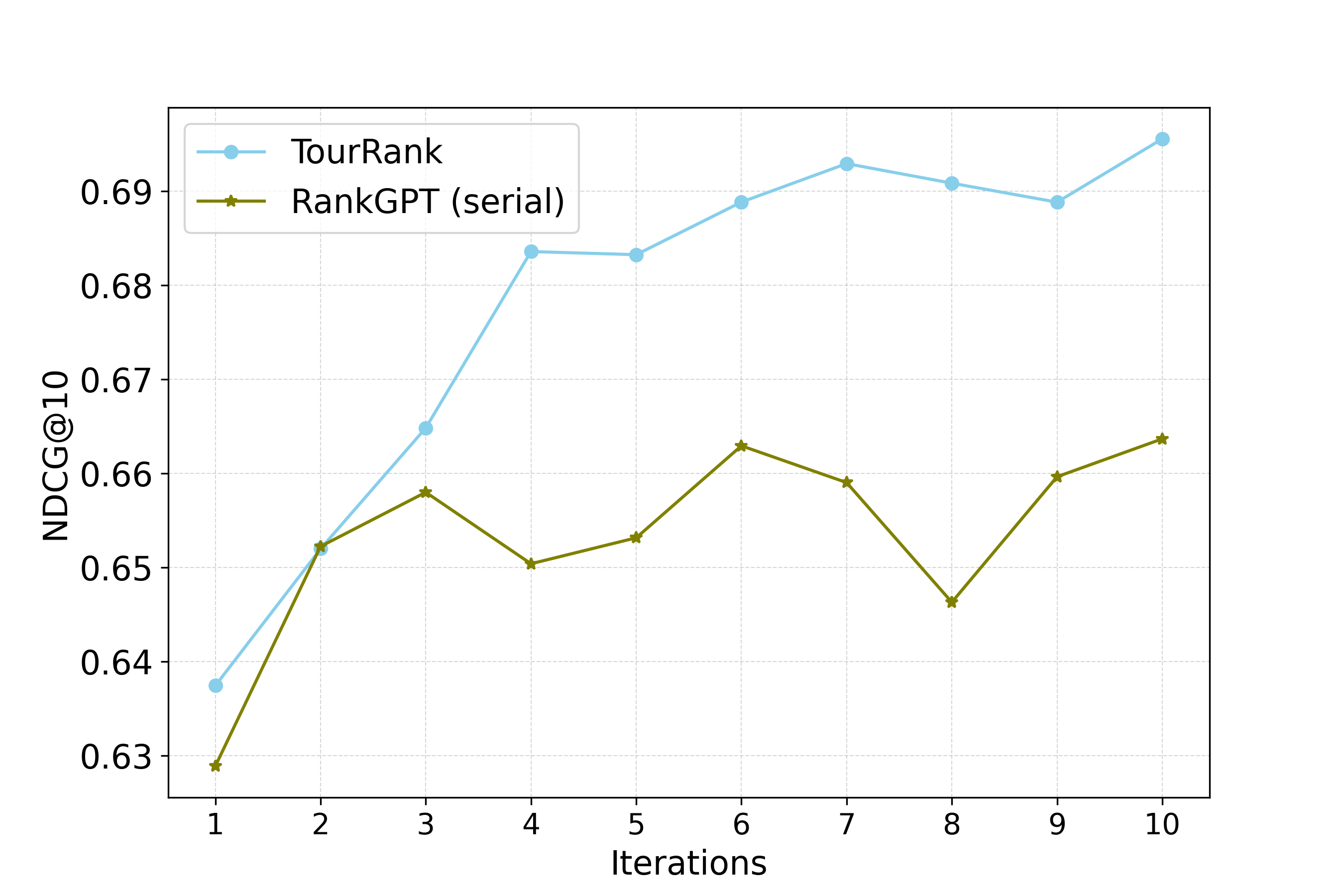}
  \caption{TREC DL 20}
  \label{chuanxing_dl20}
\end{subfigure}
\caption{The comparison of NDCG@10 between running RankGPT multiple times in serial and running TourRank-$r$ in parallel.}
\label{chuanxing}
\end{figure*}

We also run RankGPT multiple times in seriality called RankGPT (serial), that is, the documents order obtained by this iteration is used as the initial order for the next iteration. Figure \ref{chuanxing} shows the comparison of RankGPT (serial) and our TourRank. We can see that on both TREC DL 19 and TREC DL 20 datasets, the NDCG@10 of RankGPT (serial) goes up for the first three iterations, but stops going up after that. This indicates that RankGPT will reach the upper limit after a few serial runs. However, after multiple iterations (or tournaments) of TourRank-$r$, the NDCG@10 still continues to rise and performs much better than RankGPT (serial). 

RankGPT (serial) and TourRank after the same $r$ iterations: (1) The number of documents inputted to LLMs are both about $2r*N$; (2) The time complexity $O(K-1)$ of TourRank is also significantly less than $O(\frac{r}{10}*N)$ of RankGPT (serial); (3) The performance of TourRank is significantly better than RankGPT (serial). These indicate that TourRank can achieve a better balance between effectiveness and efficiency.

\section{The Detail Hyperparameters of TourRank}
\label{hyperparameters}

The detail of hyperparameters of TourRank are shown in Table \ref{Hyperparameters of TourRank}.

\begin{table*}[h] %\tiny
\centering
\resizebox{0.88\textwidth}{!}{%
\begin{tabular}{c|c|c}
\toprule
\textbf{Parameters} & \textbf{Explanation} & \textbf{Value} \\
\midrule
$R$ & \parbox{10cm}{The rounds of tournament in TourRank.} & 10 \\
\midrule
$K$ & \parbox{10cm}{One tournament contains $K-1$ times selection stages.} & 6 \\
\midrule
\multirow{6}{*}{$N_k$} & \multirow{6}{*}{\parbox{10cm}{The number of candidate documents in $k$-th selection stages in a tournament. ($k \in \{1, \cdots, K\}$)}} & $N_1 = 100$ \\
 & & $N_2 = 50$ \\
 & & $N_3 = 20$ \\
 & & $N_4 = 10$ \\
 & & $N_5 = 5$ \\
 & & $N_6 = 2$ \\
\midrule
\multirow{5}{*}{$G/n/m$} & \multirow{5}{*}{\parbox{10cm}{%
    $G$: Divide candidate documents into $G$ groups. \\
    $n$: Each group has $n$ documents. \\
    $m$: Select $m$ documents from each group.
}} & $100 \to 50: 5/20/10$ \\
 & & $50 \to 20: 5/10/4$ \\
 & & $20 \to 10: 1/20/10$ \\
 & & $10 \to 5: 1/10/5$ \\
 & & $5 \to 2: 1/5/2$ \\
\bottomrule
\end{tabular}
}
\caption{Hyperparameters of TourRank.}
\label{Hyperparameters of TourRank}
\end{table*}

Table \ref{specific points} shows the specific points of candidate document after 1 time tournament under the setting of our experiments. 

\begin{table*}[h] %\tiny %\footnotesize %\tiny \scriptsize \footnotesize \small \normalsize
\centering
\resizebox{0.33\textwidth}{!}{
\begin{tabular}{c|c}
\toprule
% \hline
\textbf{Number of Docs} & \textbf{Points of Docs} \\
\midrule
$N_6=2$ & $5$ \\
\midrule
$N_5-N_6=3$ & $4$ \\
\midrule
$N_4-N_5=5$ & $3$ \\
\midrule
$N_3-N_4=10$ & $2$ \\
\midrule
$N_2-N_3=30$ & $1$ \\
\midrule
$N_1-N_2=50$ & $0$ \\
% \hline
\bottomrule
\end{tabular}
}
\caption{The specific points of all documents after one tournament in our experimental settings.}
\label{specific points}
\end{table*}

% \clearpage
\section{Prompts}
Table \ref{The prompt of grouping stage of TourRank.} shows the prompt used in the grouping and selction stage (Figure \ref{tournament_grouping} (b)) of TourRank. 

\begin{table*}[t]
    \centering
    \begin{tcolorbox}[width=0.9\linewidth]

    \textbf{system:} You are an intelligent assistant that can compare multiple documents based on their relevancy to the given query. \\

    \noindent
    \textbf{user:} I will provide you with the given query and $n$ documents. Consider the content of all the documents comprehensively and select the $m$ documents that are most relevant to the given query: $query$. \\
    
    \noindent
    \textbf{assistant:} Okay, please provide the documents. \\
    
    \noindent
    \textbf{user:} Document 1: $Doc_1$
    
    \noindent
    \textbf{assistant:} Received Document 1. \\
    
    \noindent
    \textbf{user:} Document 2: $Doc_2$
    
    \noindent
    \textbf{assistant:} Received Document 2. \\
    
    \noindent
    (User input more documents to assistant.) \\
    
    \noindent
    \textbf{user:} The Query is: $query$. Now, you must output the top $m$ documents that are most relevant to the Query using the following format strictly, and nothing else. Don't output any explanation, just the following format: 
    
    \noindent
    Document 3, ..., Document 1
    
    \end{tcolorbox}
\caption{The prompt of the grouping and selection stage of TourRank.}
\label{The prompt of grouping stage of TourRank.}
\end{table*}

% \section{}

\begin{table*}[h] %\tiny
\centering
\resizebox{0.575\textwidth}{!}{%
\begin{tabular}{c|c|cc}
\toprule
\textbf{Methods} & \textbf{LLMs} & \textbf{TREC DL 19} & \textbf{TREC DL 20}\\
\midrule
% \midrule
BM25 & - & 50.58 & 47.96 \\
\midrule
% \midrule
% BM25 & \multirow{5}{*}{Llama 3 8B} & 50.58 & 47.96 \\
RankGPT & \multirow{5}{*}{Llama-3-8B-Instruct} & 59.48 & 54.47 \\
% TourRank-1 & & 64.92 & 62.22 \\
Setwise.heapsort & & 58.31 & 44.92 \\
Setwise.bubblesort & & 49.85 & 34.42 \\
TourRank-2 & & 71.17 & 66.84 \\
TourRank-10 & & \textbf{73.30} & \textbf{67.25} \\
\midrule
RankGPT & \multirow{5}{*}{vicuna-13b-v1.5} & 63.90 & \textbf{60.80} \\
% TourRank-1 & & 64.92 & 62.22 \\
Setwise.heapsort & & 65.90 & 58.30 \\
Setwise.bubblesort & & 62.20 & 60.20 \\
TourRank-2 & & 58.55 & 49.37 \\
TourRank-10 & & \textbf{66.21} & 59.60 \\
\midrule
RankGPT & \multirow{3}{*}{Mistral-7B-Instruct-v0.2} & 61.90 & 58.54 \\
% TourRank-1 & & 62.38 & 57.86 \\
TourRank-2 & & 65.85 & 62.31 \\
TourRank-10 & & \textbf{68.64} & \textbf{65.04} \\
\midrule
RankGPT & \multirow{3}{*}{gpt-4-turbo} & 72.67 & 69.48 \\
TourRank-1 & & 72.46 & 67.38 \\
TourRank-5 & & \textbf{74.13} & \textbf{69.79} \\
\midrule
RankGPT & \multirow{3}{*}{gpt-4o-mini} & 72.85 & \textbf{70.35} \\
TourRank-1 & & 73.35 & 67.89 \\
TourRank-5 & & \textbf{75.57} & 70.07 \\
\bottomrule
\end{tabular}
}
\caption{NDCG@10 of TourRank and RankGPT based on open-source LLMs, Mistral-7B, Llama-3-8B, vicuna-13b and OpenAI's API, gpt-4-turbo and gpt-4o-mini.}
\label{open-source LLMs}
\end{table*}

% \begin{mdframed}
% \textbf{system:} You are an intelligent assistant that can compare multiple documents based on their relevancy to the given query. \\

% \noindent
% \textbf{user:} I will provide you with the given query and $n$ documents. Consider the content of all the documents comprehensively and select the $m$ documents that are most relevant to the given query: $query$. \\

% \noindent
% \textbf{assistant:} Okay, please provide the documents. \\

% \noindent
% \textbf{user:} Document 1: The content of $Doc 1$.

% \noindent
% \textbf{assistant:} Received Document 1. \\

% \noindent
% \textbf{user:} Document 2: The content of $Doc 2$.

% \noindent
% \textbf{assistant:} Received Document 2. \\

% \noindent
% (User input more documents to assistant.) \\

% \noindent
% \textbf{user:} The Query is: $query$. Now, you must output the top $m$ documents that are most relevant to the Query using the following format strictly, and nothing else. Don't output any explanation, just the following format: 

% \noindent
% Document 3, ..., Document 1

% \end{mdframed}

% \twocolumn

\end{document}